\documentstyle[12pt,epsf]{article}
\begin{document}

\title{Nonlinear equation for curved stationary flames}

\author{Kirill~A.~Kazakov${\mbox{}}^{1,2}$\thanks{E-mail:
$Kirill.Kazakov@fysik.uu.se$} ~~and
Michael~A.~Liberman${\mbox{}}^{1,3}$\thanks{E-mail:
$Michael.Liberman@fysik.uu.se$}}

\maketitle

\begin{center}
${\mbox{}}^{1}$ Department of
Physics, Uppsala University, \\ Box 530, S-751 21, Uppsala, Sweden \\
${\mbox{}}^{2}$ Moscow State University, Physics Faculty,
Department of Theoretical Physics, \\ 117234, Moscow, Russian
Federation
\\ ${\mbox{}}^{3}$P.~Kapitsa Institute
for Physical Problems, Russian Academy of Sciences,  \\
117334, Moscow, Russian Federation
\end{center}

\begin{abstract}
A nonlinear equation describing curved stationary flames with
arbitrary gas expansion $\theta = \rho_{{\rm fuel}}/\rho_{{\rm
burnt}}$, subject to the Landau-Darrieus instability, is obtained
in a closed form without an assumption of weak nonlinearity. It is
proved that in the scope of the asymptotic expansion for $\theta
\to 1,$ the new equation gives the true solution to the problem of
stationary flame propagation with the accuracy of the sixth order
in $\theta - 1.$ In particular, it reproduces the stationary
version of the well-known Sivashinsky equation at the second order
corresponding to the approximation of zero vorticity production.
At higher orders, the new equation describes influence of the
vorticity drift behind the flame front on the front structure. Its
asymptotic expansion is carried out explicitly, and the resulting
equation is solved analytically at the third order. For arbitrary
values of $\theta,$ the highly nonlinear regime of fast flow
burning is investigated, for which case a large flame velocity
expansion of the nonlinear equation is proposed.

\end{abstract}


\unitlength=1pt

\noindent

\section{Introduction}

Curved flame propagation is one of the most important and
difficult issues in the combustion theory. Despite considerable
efforts, its closed theoretical description is still lacking.
Perhaps the main reason underlying complexity of the problem is
the intrinsic instability of zero-thickness flames, the
Landau-Darrieus (LD) instability \cite{landau,darrieus}. In view
of this, evolution of the flame front cannot be prescribed in
advance. Instead, it should be determined in the course of joint
analysis of the flow dynamics outside the flame front and the heat
conduction~--~species diffusion processes inside. In general,
nonlinear interaction of different perturbation modes under the
smoothing influence of thermal conduction leads to the formation
of a steady curved flame front configuration with the curvature
radius of the order $10 L_f,$ where $L_f$ is the flame front
thickness. This estimate for the curvature radius can be obtained
from the linear theory of the LD-instability \cite{pelce}, where
it corresponds to the cutoff wavelength for the front
perturbations; it is also confirmed by 2D numerical experiments on
the flame dynamics \cite{siv1,bychkov2}.

In essential, difficulties encountered in investigation of the
nonlinear development of the LD-instability are twofold:

1) Perturbation analysis of the nonlinearity of flame dynamics
with arbitrary gas expansion is generally inadequate. In
particular, it is completely irrelevant to the problem of
formation of the stationary flame configurations.

2) Finite vorticity production in the flame implies that the flow
dynamics downstream, and therefore the flame front dynamics
itself, is essentially nonlocal. The latter means that the
non-locality of equations governing flame propagation is more
complex than that encountered in the linear theory and described
by the Landau-Darrieus operator.

Only in the case of small gas expansion can the problem be treated
both perturbatively and locally, since then the amplitudes of
perturbations remain small compared to their wavelengths at all
stages of development of the LD-instability \cite{zel1,zel2,siv2},
and the flow is potential both up- and downstream in the lowest
order in $\theta - 1,$ where $\theta$ is the ratio of the fuel
density and the density of the burnt matter. In this
approximation, the nonlinear evolution of the front perturbations
is described by the well-known Sivashinsky equation \cite{siv2}.

The nonlinear dynamics of flames with finite gas expansion has
been the subject of a number of papers some of which are briefly
considered here as the characteristic examples of dealing with the
difficulties mentioned above.

To render the problem tractable, certain simplifying assumptions
has been introduced by various authors in attempt to weaken or
even get rid of one of the two above features inherent to the
nonlinear flame dynamics. In Ref.~\cite{frankel}, the vorticity
production in the flame is completely neglected [see the point 2)
above]. The mass conservation and the constant normal flame
velocity are taken as the conditions governing flame dynamics. The
problem is reduced thereby to the well-known electrodynamic
problem of determining the single layer potential with constant
charge distribution proportional to the gas expansion. However,
neglecting the vorticity downstream breaks the continuity of
tangential velocity components as well as the constant jump of
pressure across the flame. This model thus violates the basic
conservation laws to be satisfied across the flame front. Note
that the number of neglected degrees of freedom (three in the
3-dimensional case) just corresponds to the number of broken
conservation laws (two tangential velocity components and the
scalar pressure).

In contrast, in Ref.~\cite{bychkov1}, an attempt is made to take
into account the vorticity production in stationary flames, under
the assumption of weak nonlinearity [see the point 1) above]. From
the mathematical point of view, however, the assumptions of weak
nonlinearity and stationarity contradict each other. Using them
simultaneously turns out to be inconsistent except for the case of
small gas expansion. Indeed, let us consider a weakly curved flame
front propagating in $z$-direction with unit normal
velocity\footnote{It is convenient to use dimensionless velocity
normalized on the velocity of a plane flame, $U_f.$} with respect
to an initially uniform fuel; the transverse coordinates will be
denoted {\bf x}. It is not difficult to show that in this case the
flow equations up- and downstream together with the conservation
laws at the flame front imply the following relation between the
flame front position, $z = f({\bf x },t) - t,$ and $z$-component
of the fuel velocity ${\bf v} = {\bf v}({\bf x},z,t)$ just ahead
of the front, $u_{-} = {\rm v}_z ({\bf x},f({\bf x},t)-0,t),$
\begin{eqnarray}\label{rel1}
u_{-} + \frac{\theta + 1}{2\theta}\hat{\Phi}^{-1}\frac{\partial
u_{-}}{\partial t} = \frac{\theta - 1}{2}\hat{\Phi}f,
\end{eqnarray}
\noindent where $\hat{\Phi}$ denotes the Landau-Darrieus operator
defined by
$$(\hat{\Phi}f)({\bf x}) = \int_{- \infty}^{+ \infty} d{\bf k} |{\bf k}|
f_{\bf k} \exp (i {\bf k x}),$$ $f_{\bf k}$ being the Fourier
transform of $f.$ (It is assumed here that ${\bf v}(z = - \infty)
= 0$ in the laboratory frame of reference. Details can be found,
e.g., in Ref.~\cite{bychkov3}) On the other hand, to the leading
order in the front slope $\partial f/\partial{\bf x },$ the
condition of unit normal flame velocity with respect to the fuel
gives
\begin{eqnarray}\label{rel2}
u_{-}\ = \frac{\partial f}{\partial t} +
\frac{1}{2}\left(\frac{\partial f}{\partial {\bf x}}\right)^2.
\end{eqnarray}
\noindent In the regime of steady flame propagation,
$$\frac{\partial u_{-}}{\partial t} = 0, ~~\frac{\partial f}{\partial t} = - W,$$
where $W\sim (\partial f/\partial{\bf x})^2$ is the flame velocity
increase due to the front curvature. Substituting this into
relations (\ref{rel1}), (\ref{rel2}) gives
\begin{eqnarray}\label{rel3}
- W + \frac{1}{2}\left(\frac{\partial f}{\partial {\bf
x}}\right)^2 = \frac{\theta - 1}{2}\hat{\Phi}f.
\end{eqnarray}
\noindent Since the right hand side of this equation is
linear\footnote{See the Appendix.} in the slope, while the left
hand side is only quadratic, nonlinearity can be considered small
only if $\theta - 1$ is small, in which case one has $\partial
f/\partial {\bf x} \sim \theta - 1,$ $W \sim (\theta - 1)^2.$
Thus, for arbitrary $\theta,$ the weak nonlinearity approach to
the stationary flames, advocated in \cite{bychkov1}, turns out to
be self-contradictory. From the physical point of view, this means
that a weakly curved flame with finite gas expansion cannot be
stationary. Instead, equations (\ref{rel1}), (\ref{rel2}) should
be considered in this case as determining evolution of the small
perturbations in time. Indeed, substituting Eq.~(\ref{rel2}) into
Eq.~(\ref{rel1}), we obtain, in the linear approximation,
\begin{eqnarray}\label{rel4}
\frac{\theta + 1}{2\theta}\hat{\Phi}^{-1}\frac{\partial^2
f}{\partial t^2} + \frac{\partial f}{\partial t} - \frac{\theta -
1}{2}\hat{\Phi}f = 0,
\end{eqnarray}
\noindent which is nothing but the well-known LD dispersion
relation for the perturbation growth rate, written in the
coordinate space \cite{landau,darrieus}.

In practice, the discrepancy in the orders of magnitude, expressed
by Eq.~(\ref{rel3}), shows itself as the impossibility to
correctly develop perturbation expansion in powers of the slope.
For instance, the jump of the pressure field across the flame
front, found in Ref.~\cite{bychkov1}, Eq.~(33), turns out to be
velocity-dependent already for zero thickness flames, while
according to the well-known exact result \cite{matalon} this jump
is constant along the flame front, and is equal to $1 - \theta.$

It should be stressed that the above-mentioned inconsistency
resides in the equations of motion themselves; it is inherent,
therefore, to the flame propagation in tubes as well as to
unbounded flames. Indeed, even in the case of very narrow tubes,
numerical experiments on flames with $\theta = 6~\div~8$ give
values of about $1.5~\div~2.0$ for the slope \cite{bychkov2}.

Finally, we mention Ref.~\cite{zhdanov}, where a non-stationary
equation describing the interaction of perturbations at the early
stage of development of LD instability is obtained at the second
order of nonlinearity, under a certain model assumption concerning
the structure of downstream flow. Namely, it is assumed that there
exists a {\it local} relation between the pressure field and the
potential mode of the velocity downstream (expressed by the
Bernoulli equation). However, as we mentioned in 2), the flow
dynamics downstream is essentially nonlocal, in particular, the
pressure field is expressed through the velocity field by an
integral relation, its kernel being an appropriate Green function
of the Laplace operator. From the work \cite{zhdanov}, one can see
once again that the perturbative treatment of nonlinearities is
not applicable to the stationary flames with arbitrary $\theta:$
since, at the first order, the time derivatives of the front
position are linearly related to its space derivatives through
Eq.~(\ref{rel4}), stationary form of the non-stationary
Zhdanov-Trubnikov equation depends on the way this relation is
used before the time derivatives are omitted. Transition to the
stationary regime in this equation is therefore ambiguous.

In view of what has been said, we arrive at the conclusion that
the stationary flames can only be treated consistently in the
framework of the small $\theta - 1$ expansion. Thus, the problem
of describing the stationary flames, and more generally,
non-stationary flames in the fully developed nonlinear regime, can
be formulated as the problem of deriving an expansion of unknown
exact equation, governing flame dynamics, in powers of $\theta -
1.$

The purpose of the present paper is to show that in the stationary
case, such an expansion can indeed be developed beyond the second
order, the order of validity of the Sivashinsky equation. We found
it convenient to employ the model assumption of
Ref.~\cite{zhdanov} in our analysis, which allows one to obtain an
equation describing flames with arbitrary gas expansion in a
closed form, without the assumption of weak nonlinearity. This
will be shown in Sec.~\ref{deriv} on the basis of simple geometric
considerations. That the equation obtained actually provides the
true expansion of (unknown) exact equation with the accuracy of
the sixth(!) order in $\theta - 1$ is proved in
Sec.~\ref{estimate}. This expansion is carried out explicitly in
Sec.~\ref{4order}. The third order equation turns out to have the
functional structure analogous to the Sivashinsky equation, and
therefore can be solved analytically. This solution is found in
Sec.~\ref{3order}. Finally, the model equation obtained in
Sec.~\ref{deriv} is invoked in Sec.~\ref{largev} for investigation
of the highly nonlinear regime of fast flow burning, where a large
flame velocity expansion of this equation is proposed. The results
of the present work are summarized in Sec.~\ref{conclud}. Some
mathematical results used in the text are derived in the Appendix.

\section{Flow equations and conservation laws}\label{general}

Let us consider a 2D flame propagating in an initially uniform
premixed fluid. Since our main concern is the influence of finite
gas expansion on the flame structure, we will assume in what
follows that the fluid is an ideal gas with constant thermal
conduction and constant specific heat. The ratio of the heat and
mass diffusivities (the Lewis number) is taken to be unity. For
definiteness, we consider flame propagation in a tube of arbitrary
width with ideally slip adiabatic walls. Under the assumption that
development of the LD instability ends up with the formation of a
steady flame configuration, we go over to the reference frame of
the stationary flame. Let the Cartesian coordinates $(x,z)$ be
chosen so that $z$-axis is parallel to the tube walls, $z = -
\infty$ being in the fresh fuel. It will be convenient to
introduce the following dimensionless variables
$$(\eta;\xi )=(x/R;z/R), \  (w;u) =({\rm v}_{x}/U_{f};{\rm v}_{z}/U_{f}),$$
$$\Pi =(P-{P}_{f}) /{\rho }_{-}{U_f}^{2},$$
where $U_{f}$ is the velocity of a plane flame front, $P_f$ is the
initial pressure in the fuel far ahead of the flame front, and $R$
is some characteristic length of the problem (e.g., the cut-off
wavelength). The fluid density will be normalized on the fuel
density $\rho_{-}.$ As always, we assume that the process of flame
propagation is nearly isobaric. Then the velocity and pressure
fields obey the following equations in the bulk
\begin{eqnarray}\label{flow1}&&
\frac{\partial u}{\partial\xi} + \frac{\partial w}{\partial\eta} = 0,
\end{eqnarray}
\begin{eqnarray}\label{flow2}&&
u \frac{\partial u}{\partial\xi} + w \frac{\partial u}{\partial\eta}
= - \frac{1}{\rho}\frac{\partial \Pi}{\partial\xi},
\end{eqnarray}
\begin{eqnarray}\label{flow3}&&
u \frac{\partial w}{\partial\xi} + w \frac{\partial
w}{\partial\eta} = - \frac{1}{\rho}\frac{\partial
\Pi}{\partial\eta}.
\end{eqnarray}
\noindent

The above flow equations are complemented by the following
conservation laws to be satisfied across the flame front
\begin{eqnarray}\label{conserv1}&&
u_{+} - \theta u_{-} - f'(w_{+} - \theta w_{-}) = 0,
\end{eqnarray}
\begin{eqnarray}\label{conserv2}&&
w_{+} - w_{-} + f'(u_{+} - u_{-}) = 0,
\end{eqnarray}
\begin{eqnarray}\label{conserv3}&&
\Pi_{+} - \Pi_{-} = - (\theta - 1),
\end{eqnarray}
\noindent where the flame front position is given by $\xi =
f(\eta),$ the subscripts "$+$" and "$-$" mean that the
corresponding quantity should be evaluated at $\xi = f(\eta) + 0$
and $\xi = f(\eta) - 0,$ respectively, and
$$N\equiv \sqrt{1 + f'^2},~~f'\equiv \frac{d f}{d \eta}.$$

Finally, the normal velocity of the fuel at the flame front is
unity, which is expressed in the form of the evolution equation
\begin{eqnarray}\label{evolution}&&
u_{-} - f'w_{-} = N.
\end{eqnarray}
\noindent Equations (\ref{conserv1})--(\ref{evolution}) are
written for zero thickness flames. These are of primary interest,
since in the majority of cases the thickness of the flame front is
small compared with the fluid-dynamical scale of the problem.
Formal generalization to the case of small but nonzero thickness
is straightforward and will be done in Sec.~\ref{account}.

In principle, the system of equations (\ref{flow1}) --
(\ref{evolution}) completely determines stationary flame
configuration as well as the flows upstream and downstream. To
derive an equation for the flame front function $f(\eta),$ one has
to find solution of the bulk equations (\ref{flow1}) --
(\ref{flow3}) for the fuel and combustion products with
appropriate boundary conditions on the walls, subject to the
conservation laws (\ref{conserv1}) -- (\ref{conserv3}) which are a
kind of boundary conditions on the flame front. The requirement
that the obtained solution satisfies Eq.~(\ref{evolution}) then
gives an equation for the function $f(\eta)$ itself.

As to the flow upstream, the corresponding solution is readily
obtained: since the flow is potential at $\xi = - \infty$ ($u =
V,$ $w = 0,$ where $V$ is the velocity of the flame in the rest
frame of reference of the fuel), it is potential for every
$\xi<f(\eta)$ in view of the Helmholtz theorem \cite{landafshitz},
thus
\begin{eqnarray}&&\label{solup1}
u \equiv V + \tilde{u} = V + \int\limits_{- \infty}^{+ \infty}d k
~\tilde{u}_k \exp (|k|\xi + i k\eta),
\end{eqnarray}
\begin{eqnarray}&&\label{solup2}
w = \hat{H}\tilde{u},
\end{eqnarray}
\begin{eqnarray}&&\label{solup3}
\Pi + \frac{1}{2}(u^2 + w^2) = {\rm const_1},
\end{eqnarray}
\noindent
where the Hilbert operator
$$(\hat{H}f)(\eta) = \frac{1}{\pi}{\rm p.v.}\int\limits_{- \infty}^{+ \infty}
d\zeta\frac{f(\zeta)}{\zeta - \eta},$$ "${\rm p.v.}$" denoting the
principal value. Although the relation $w = \hat{H}\tilde{u}$
between the velocity components upstream is nonlocal, it is
expressed in terms of the transverse coordinate $\eta$ only.

Things become more complicated downstream. There, no relation
exists for the variables $u, ~w, ~\Pi,$ which can be expressed in
terms of $\eta$ alone, since the $\xi$-dependence of these
variables is unknown because of the presence of vorticity produced
by the curved flame. Nevertheless, we will assume following
Ref.~\cite{zhdanov} that a potential mode~${\bf v}_{p} =
(w_p,u_p)$ can be extracted from the downstream velocity~${\bf v}
= (w,u),$ such that the following Bernoulli-type relation holds
between ${\bf v}_{p}$ and~$\Pi$
\begin{eqnarray}&&\label{model}
\Pi + \frac{1}{2\theta}(u_p^2 + w_p^2) = {\rm const_2}.
\end{eqnarray}
\noindent
Then Eqs.~(\ref{flow1}) -- (\ref{flow3}) can be rewritten as
\begin{eqnarray}\label{continuity}&&
\frac{\partial u}{\partial\xi} + \frac{\partial w}{\partial\eta} =
0, ~~\frac{\partial u_p}{\partial\xi} + \frac{\partial
w_p}{\partial\eta} = 0,
\end{eqnarray}
\begin{eqnarray}\label{flowdown1}&&
u \frac{\partial u}{\partial\xi} + w \frac{\partial u}{\partial\eta}
- u_p \frac{\partial u_p}{\partial\xi} - w_p \frac{\partial u_p}{\partial\eta}
= 0,
\end{eqnarray}
\begin{eqnarray}\label{flowdown2}&&
u \frac{\partial w}{\partial\xi} + w \frac{\partial w}{\partial\eta}
- u_p \frac{\partial w_p}{\partial\xi} - w_p \frac{\partial w_p}{\partial\eta}
= 0.
\end{eqnarray}
The general solution for the potential mode can be written
analogously to Eqs.~(\ref{solup1}), (\ref{solup2})
\begin{eqnarray}&&\label{soldown1}
u_p \equiv \theta V + \tilde{u}_p = \theta V
+ \int\limits_{- \infty}^{+ \infty}d k~\tilde{u}_k \exp (- |k|\xi + i k\eta),
\end{eqnarray}
\begin{eqnarray}&&\label{soldown2}
w_p = - \hat{H}\tilde{u}_p.
\end{eqnarray}
\noindent The model relation (\ref{model}) does not uniquely
define the potential mode, since a constant term in the
$\xi$-component of the velocity can be assigned either to the
potential mode or to the vorticity mode. As will be shown in
Sec.~\ref{estimate}, the vorticity produced in a flame is only of
the fourth order in $(\theta - 1 )$ as $\theta \to 1.$ The choice
$\theta V$ of the constant term in Eq.~(\ref{soldown1}) is fixed
therefore up to the second\footnote{One might think that since the
vorticity is of the fourth order in $\theta - 1,$ the flow is to
be potential up to the third order. One should remember, however,
that the expression of the velocity field through the vorticity
field is nonlocal. Specifically, we will see in
Sec.~\ref{estimate} that the vorticity produced in the flame can
be expressed as the $\eta$-derivative of a certain function of the
velocity, while differentiation along the flame front brings in an
extra power of $\theta - 1.$} order by the requirement that the
mass flow at $\xi = + \infty,$ where $u = u_p = {\rm const}, w =
0,$ equals that at $\xi = - \infty$.

Now, using the continuity equations (\ref{continuity}) for ${\bf
v}$ and ${\bf v}_p,$ Eq.~(\ref{flowdown1}) can be written in the
form in which $\xi$-dependence is implicit
\begin{eqnarray}\label{flowdownnew}&&
u\frac{\partial w}{\partial\eta} - w\frac{\partial
u}{\partial\eta} - u_p\frac{\partial w_p}{\partial\eta} +
w_p\frac{\partial u_p}{\partial\eta} = 0.
\end{eqnarray}
\noindent It will be shown in the next section, the above
equations for the potential mode upstream and downstream, equation
(\ref{flowdownnew}), and the conservation laws at the flame front
constitute the system of equations sufficient to derive an
equation for the function $f(\eta)$ in a closed form.

\section{Nonlinear equation for the flame front}

\subsection{Derivation}\label{deriv}

It will be shown presently that the set of equations
(\ref{conserv1}) -- (\ref{evolution}), (\ref{solup2}) --
(\ref{model}), (\ref{soldown2}), and (\ref{flowdownnew}) can be
transformed into one equation for the function $f(\eta).$ The fact
that this set is written in a form that does not explicitly
operates with the $\xi$-dependence of the flow variables makes it
unnecessary to follow the program outlined in the preceding
section to obtain an equation for $f(\eta).$ Namely, the specific
structure of the up- and downstream flows in the bulk is now
irrelevant. In particular, knowledge of the $\xi$-dependence of
the velocity and pressure fields turns out to be superfluous.
Roughly speaking, the $\xi$-dependence of a function $F(\xi,\eta)$
describing the shape of the flame front is known in advance, since
the equation $F(\xi,\eta) = 0$ can always be brought into the form
$\xi - f(\eta) = 0$ (with $f$ many-valued, in general).
Determination of the $\eta$-dependence alone of the functions
involved is therefore sufficient for the purpose of description of
the flame front structure.

In what follows, it will be convenient to introduce separate
designations for the up- and downstream velocity and pressure
fields. Namely, they will be distinguished by the {\it
super}scripts $"+"$ and $"-"$, respectively. Then, setting $\xi =
f(\eta),$ equations~(\ref{solup2}), (\ref{solup3}), (\ref{model}),
(\ref{soldown2}), and (\ref{flowdownnew}), together with the
conservation laws (\ref{conserv1}) -- (\ref{conserv3}) and the
evolution equation (\ref{evolution}), can be rewritten identically
as follows
$$\left(
\begin{array}{rcc}
u^{+} - \theta u^{-} - f'(w^{+} - \theta w^{-})&=&0\\
w^{+} - w^{-} + f'(u^{+} - u^{-})&=&0\\
\Pi^{+} - \Pi^{-}&=& - (\theta - 1)\\
u^{-} - f'w^{-}&=& N\\
w^{-}&=&\hat{H}\tilde{u}^{-}\\
\Pi^{-} + \frac{1}{2}\{(u^{-})^2 + (w^{-})^2\}&=&{\rm const_1}\\
\Pi^{+} + \frac{1}{2\theta}\{(u_p^{+})^2 + (w_p^{+})^2\}&=&{\rm
const_2}\\
w_p^{+} &=& - \hat{H}\tilde{u}_p^{+}\\
u^{+} \frac{\partial w^{+}}{\partial\eta} - w^{+}\frac{\partial
u^{+}}{\partial\eta} - u_p^{+} \frac{\partial
w_p^{+}}{\partial\eta} + w_p^{+} \frac{\partial
u_p^{+}}{\partial\eta} &=& 0
\end{array}
\right)_{\xi = f(\eta)}\eqno (*)$$

Suppose we have found a solution $f = f(\eta),$ ${\bf v}^{-} =
{\bf v}^{-}(\xi,\eta),$ ${\bf v}^{+} = {\bf v}^{+}(\xi,\eta),$
etc. of the set of equations in the large brackets in ($*$). Then,
in particular, these equations are satisfied for $\xi = f(\eta).$
On the other hand, since no operation involving $\xi$ appears in
these equations,\footnote{No such operation can appear in the
boundary conditions to these equations neither. Otherwise, steady
flame propagation would be impossible.} the function $f(\eta)$ is
one and the same for all solutions. Furthermore, since $f(\eta)$
is $\xi$-independent, it is convenient to work with the particular
solution in which all the other functions are also
$\xi$-independent, and to omit the large brackets in ($*$).
Therefore, we replace the above set of equations by the following
\begin{eqnarray}
\upsilon^{+} - \theta \upsilon^{-}
- f'(\omega^{+} - \theta \omega^{-})&=&0 \label{conservn1}\\
\omega^{+} - \omega^{-} + f'(\upsilon^{+}
- \upsilon^{-})&=&0 \label{conservn2}\\
\pi^{+} - \pi^{-}&=& - (\theta - 1) \label{ampl1}\\
\upsilon^{-} - f'\omega^{-}&=& N \label{evolutionn}\\
\omega^{-}&=&\hat{H}\tilde{\upsilon}^{-} \label{solup2n}\\
\pi^{-} + \frac{1}{2}\{(\upsilon^{-})^2
+ (\omega^{-})^2\}&=&{\rm const_1} \label{ampl2}\\
\pi^{+} + \frac{1}{2\theta}\{(\upsilon_p^{+})^2
+ (\omega_p^{+})^2\}&=&{\rm const_2} \label{ampl3}\\
\omega_p^{+} &=& - \hat{H}\tilde{\upsilon}_p^{+} \label{consist}\\
\upsilon^{+} \frac{d\omega^{+}}{d\eta} -
\omega^{+}\frac{d\upsilon^{+}}{d\eta} - \upsilon_p^{+} \frac{d
\omega_p^{+}}{d\eta} + \omega_p^{+} \frac{d\upsilon_p^{+}}{d\eta}
&=& 0, \label{phase}
\end{eqnarray}
\noindent
where $\upsilon, \omega,$ and $\pi$ are the
$\xi$-independent counterparts of the flow variables $u, w,$ and
$\Pi,$ respectively, and
\begin{eqnarray}\label{decompose}
\upsilon^{-} \equiv V + \tilde{\upsilon}^{-},
~~\upsilon_p^{+}
\equiv \theta V + \tilde{\upsilon}_p^{+}.
\end{eqnarray}
\noindent The fact that now the function $f(\eta)$ does not enter
the arguments of these variables allows us to avoid expanding them
in powers of $f$ (employed, e.g., in
Refs.~\cite{zhdanov,bychkov1}). In fact, such an expansion is
irrelevant to the issue whatever regime (stationary or not) is
considered, since all the equations governing flame propagation
are invariant with respect to the space translations, and
therefore, all terms containing powers of undifferentiated $f$
should appear in invariant combinations in the final equation for
$f.$ On the other hand, in view of the above-mentioned translation
invariance, the function $f$ itself does not need to be small even
if the front is only weakly curved (e.g., at the early stage of
development of the LD-instability). We thus see that the
$f$-dependence of the flow variables through their arguments must
eventually cancel in some way.

Let us now turn to the derivation of the equation for the function
$f(\eta).$

From the geometric point of view, Eqs.~(\ref{ampl1}),
(\ref{ampl2}), (\ref{ampl3}) determine the amplitude $\Omega$ of the
complex function
\begin{eqnarray}\label{complex}
\upsilon^{+}_p + i\omega^{+}_p \equiv \Omega e^{i\phi},
\end{eqnarray}
\noindent
while Eq.~(\ref{phase}) gives the rate of change of its phase $\phi$
along the flame front. Indeed, Eq.~(\ref{phase}) can be rewritten as
\begin{eqnarray}\label{phase1}
\frac{d\phi}{d\eta} = \frac{\upsilon^{+} d\omega^{+}/d\eta -
\omega^{+} d\upsilon^{+}/d\eta}{(\upsilon_p^{+})^2 +
(\omega_p^{+})^2}.
\end{eqnarray}
\noindent
Then Eqs.~(\ref{conservn1}), (\ref{conservn2}), (\ref{evolutionn}),
and (\ref{solup2n}) allow one to express the right hand side of
Eq.~(\ref{phase1}) in terms of the function $f(\eta),$ while
Eq.~(\ref{consist}) plays the role of the consistency condition
which gives the equation for the function $f(\eta)$ itself.

Specifically, Eqs.~(\ref{ampl1}), (\ref{ampl2}), and (\ref{ampl3})
imply that
\begin{eqnarray}\label{ampl4}
\Omega^2 = (\upsilon_p^{+})^2 + (\omega_p^{+})^2
= \theta\{(\upsilon^{-})^2 + (\omega^{-})^2\} + C,
\end{eqnarray}
\noindent with some constant $C.$ As in Eq.~(\ref{soldown1})
above, this constant is fixed up to the second order in $\theta -
1,$ since to this order the flow is potential downstream,
$\upsilon^{+} = \upsilon_p^{+},$ $\omega^{+} = \omega_p^{+}.$ The
actual value of $C$ can be found calculating Eq.~(\ref{ampl4}) at
some particular point at the flame front. At the tube walls, e.g.,
one has $\upsilon^{-} = 1,$ $\upsilon^{+} = \theta,$ $\omega^{-} =
\omega^{+} = 0,$ therefore,
\begin{eqnarray}\label{constant}
C = \theta (\theta - 1).
\end{eqnarray}
\noindent We will see in Sec.~\ref{largev} that from the point of
view of the large flame velocity expansion, another choice of the
constant $C$ is more appropriate: $C = \theta (\theta - 1) V^2.$
This differs from that of Eq.~(\ref{constant}) only in the third
order in $\theta - 1,$ since $V = 1 + O((\theta - 1)^2)$ (see
Sec.~\ref{estimate} below).

Next, solving Eqs.~(\ref{conservn1}), (\ref{conservn2}) with
respect to $\upsilon^{+},$ $\omega^{+},$ and using
Eqs.~(\ref{ampl4}), (\ref{constant}), we derive the following
expression for the phase derivative
\begin{eqnarray}\label{phase2}&&
\phi' = \frac{\left(V + \tilde{\upsilon}^{-} + \frac{\theta -
1}{N}\right) \left(\hat{H}\tilde{\upsilon}^{-} - f'\frac{\theta -
1}{N}\right)' - \left(\hat{H}\tilde{\upsilon}^{-} - f'\frac{\theta
- 1}{N}\right)\left(\tilde{\upsilon}^{-} + \frac{\theta -
1}{N}\right)'} {\theta\left[ (V + \tilde{\upsilon}^{-})^2 +
(\hat{H}\tilde{\upsilon}^{-})^2 + (\theta - 1)\right]}.
\end{eqnarray}
\noindent Eqs.~(\ref{evolutionn}) and (\ref{solup2n}) imply that
the function $\tilde{\upsilon}^{-} = \tilde{\upsilon}^{-}(\eta)$
entering this equation can be represented in terms of $f(\eta)$ by
formally inverting the operator $(1 - f'\hat{H}):$
\begin{eqnarray}\label{expansion}&&
\tilde{\upsilon}^{-} = (1 - f'\hat{H})^{-1}(N - V).
\end{eqnarray}
The right hand side of this equation can be represented as a
formal series in powers of the operator $f'\hat{H}:$
$$(1 -
f'\hat{H})^{-1}(N - V) = (1 + f'\hat{H} + f'\hat{H}f'\hat{H} +
\cdot\cdot\cdot)(N - V).$$

Now, to obtain the equation for $f(\eta),$ we have to use the
remaining equation (\ref{consist}). In terms of the quantities
$\Omega$ and $\phi,$ it takes the form
\begin{eqnarray}\label{consist2}&&
({\rm Im} + {\rm Re}\hat{H})(\Omega e^{i\phi} - \theta V) = 0,
\end{eqnarray}
\noindent where ${\rm Re} F$ and ${\rm Im} F$ denote the real and
imaginary parts of the complex function $F,$ respectively, and we
used the fact that the Hilbert operator is real (see the
Appendix). To combine Eqs.~(\ref{ampl4}), (\ref{phase2}), and
(\ref{consist2}) into one, one has to solve the latter with
respect to $\phi.$ This can be done as follows.

Noting the relations ${\rm Im} i F = {\rm Re} F,$
${\rm Re} i F = - {\rm Im} F,$ we can rewrite Eq.~(\ref{consist2})
identically as
\begin{eqnarray}\label{consist3}&&
{\rm Im}(1 + i\hat{H})(\Omega e^{i\phi} - \theta V) = 0,
\end{eqnarray}
\noindent
or as
\begin{eqnarray}\label{consist4}&&
{\rm Re}(i - \hat{H})(\Omega e^{i\phi} - \theta V) = 0.
\end{eqnarray}
\noindent Acting by the operator $\hat{H}$ on Eq.~(\ref{consist4})
from the left, and using the identity (\ref{h1}) of the Appendix,
this equation can be rewritten also as
\begin{eqnarray}\label{consist5}&&
{\rm Re}(i\hat{H} + 1)(\Omega e^{i\phi} - \theta V) = 0.
\end{eqnarray}
\noindent
Together, Eqs.~(\ref{consist3}) and (\ref{consist5}) imply that
\begin{eqnarray}\label{consist6}&&
(1 + i\hat{H})(\Omega e^{i\phi} - \theta V) = 0.
\end{eqnarray}
\noindent Let us consider the structure of Eq.~(\ref{consist6})
more closely.

{\it Lemma:} All solutions of the equation
\begin{eqnarray}\label{ring1}&&
(1 + i\hat{H}) X = 0
\end{eqnarray}
\noindent span the ring.

{\it Proof:} First, let us show that if $X$ is a solution of
Eq.~(\ref{ring1}), then its square $X^2$ also is. Indeed, using
the identity (\ref{h8}) of the Appendix, one has
\begin{eqnarray}&&
2 i \hat{H}X^2 = 2\hat{H}(X\hat{H}X) = (\hat{H}X)^2 - X^2 = - X^2
- X^2,\nonumber
\end{eqnarray}
\noindent therefore,
\begin{eqnarray}&&
(1 + i\hat{H}) X^2 = 0.
\nonumber
\end{eqnarray}
\noindent In view of the linearity of the Hilbert operator, the
sum $X_1 + X_2$ of two solutions $X_1,$ $X_2$ of Eq.~(\ref{ring1})
also is a solution,
\begin{eqnarray}\label{ring2}&&
(1 + i\hat{H}) (X_1 + X_2) = 0,
\end{eqnarray}
\noindent and so are the squares $X_1^2,$ $X_2^2,$ and $(X_1 +
X_2)^2,$ as we just proved. We have, therefore,
\begin{eqnarray}&&
0 = (1 + i\hat{H}) (X_1 + X_2)^2 = (1 + i\hat{H}) (X_1^2 + 2 X_1
X_2 + X_2^2) = 2 (1 + i\hat{H})(X_1 X_2), \nonumber
\end{eqnarray}
\noindent or
\begin{eqnarray}\label{ring3}&&
(1 + i\hat{H})(X_1 X_2) = 0.
\end{eqnarray}
\noindent In particular, it follows from Eq.~(\ref{ring3}) by
induction that for any solution $X$ of Eq.~(\ref{ring1}) and any
positive integer $n,$
\begin{eqnarray}&&
(1 + i\hat{H}) X^n = 0, ~n\in N. \nonumber
\end{eqnarray}
\noindent In mathematical terminology, the properties
(\ref{ring2}) and (\ref{ring3}) reveal the ring structure of
solutions of Eq.~(\ref{ring1}).

This result can be used to solve Eq.~(\ref{consist6}) with respect to $\phi.$
Namely, assuming that $|\Omega e^{i\phi}/\theta V - 1|<1,$ and taking the
infinite sum of the powers
$$\left(\frac{\Omega e^{i\phi}}{\theta V} - 1\right)^n$$
multiplied by the factors $\frac{1}{n},$ we obtain
\begin{eqnarray}&&
(1 + i\hat{H})\sum\limits_{n = 1}^{\infty}
\frac{1}{n}\left(\frac{\Omega e^{i\phi}}{\theta V} - 1\right)^n =
(1 + i\hat{H})\ln\left\{1 + \left(\frac{\Omega e^{i\phi}}{\theta
V} - 1\right)\right\}
\nonumber\\&&
= (1 + i\hat{H})\left(\ln\frac{\Omega}{\theta V} + i\phi \right) = 0,
\end{eqnarray}
\noindent
or
\begin{eqnarray}&&\label{consist7}
\phi = - \hat{H}\ln\frac{\Omega}{\theta V}.
\end{eqnarray}
\noindent This solution is then analytically continued for all
values of $\Omega,$ $\phi.$

With the help of Eqs.~(\ref{ampl4}) -- (\ref{phase2}), and
(\ref{consist7}) we can now write the equation for the function
$f(\eta)$ we are looking for
\begin{eqnarray}\label{main}&&
\frac{\theta}{2}\frac{d}{d\eta}\ln\left\{(V + \upsilon)^2 +
(\hat{H}\upsilon)^2 + (\theta - 1)\right\} \nonumber\\&& =
\hat{H}\left\{\frac{\left(V + \upsilon + \frac{\theta -
1}{N}\right) \left(\hat{H}\upsilon - f'\frac{\theta -
1}{N}\right)' - \left(\hat{H}\upsilon - f'\frac{\theta -
1}{N}\right)\left(\upsilon + \frac{\theta - 1}{N}\right)'}{(V +
\upsilon)^2 + (\hat{H}\upsilon)^2 + (\theta - 1)}\right\},
\end{eqnarray}
\noindent with the denotation
\begin{eqnarray}\label{expansion1}&&
\upsilon \equiv (1 - f'\hat{H})^{-1}(N - V).
\end{eqnarray}

Before we proceed to the investigation of this equation, let us
consider the question of its accuracy.

\subsection{Accuracy assessment}\label{estimate}

We derived Eq.~(\ref{main}) using only exact transformations of
the equations (\ref{conservn1}) -- (\ref{phase}). Therefore, its
accuracy is determined entirely by the accuracy of the underlying
model assumption expressed by Eq.~(\ref{model}), which can be
estimated as follows.

As we have mentioned above, the exact equation for the stationary
flame front can, in principle, be obtained from the system of
equations (\ref{flow1}) -- (\ref{evolution}), of which only the
$\eta$-component of the Euler equations, Eq.~(\ref{flow3}), is not
present in the set ($*$) of equations describing our model. The
question of the model accuracy, therefore, is the question of the
accuracy to which Eq.~(\ref{flow3}) is satisfyed by the model
solution. To answer the latter, we need an explicit expression for
the vorticity produced in the flame, which will be derived
presently.

It was shown in Ref.~\cite{matalon} that with the help of the
conservation laws (\ref{conserv1}) -- (\ref{conserv3}) and the
evolution equation (\ref{evolution}), the value of the vorticity
just behind the flame front can be explicitly expressed in terms
of the fuel velocity. Namely, with the help of Eqs.~(5.32) and
(6.15) of the work \cite{matalon}, the jump $\{\sigma\}_{-}^{+}$
of the vorticity
$$\sigma = \frac{\partial u}{\partial\eta}
- \frac{\partial w}{\partial\xi}$$ across the flame front can be written,
in the 2D stationary case, as
\begin{eqnarray}\label{vort1}&&
\{\sigma\}_{-}^{+} = - \frac{\theta - 1}{\theta N}
\left(\hat{D}w_{-} + f'\hat{D}u_{-} + \frac{1}{N}\hat{D}f'\right),
\end{eqnarray}
\noindent
where
\begin{eqnarray}\label{operator}
\hat{D}\equiv \left(w_{-} + \frac{f'}{N}\right)\frac{d}{d\eta}.
\end{eqnarray}
Differentiating the evolution equation (\ref{evolution}) and
writing Eq.~(\ref{vort1}) longhand, expression in the brackets can
be represented as a total derivative
\begin{eqnarray}\label{vort2}&&
\hat{D}w_{-} + f'\hat{D}u_{-} + \frac{1}{N}\hat{D}f'\equiv
w_{-}'w_{-} + \frac{(f'w_{-})'}{N} + \frac{(f')^2 u_{-}'}{N} +
f'u_{-}'w_{-} + \frac{N'}{N} \nonumber\\&&
 = \frac{(w_{-}^2)'}{2} + \frac{(u_{-} - N)'}{N}
+ \frac{(N^2 - 1) u_{-}'}{N} + u_{-}'(u_{-} - N) + \frac{N'}{N} =
\frac{(u_{-}^2 + w_{-}^2)'}{2}.
\end{eqnarray}
\noindent Since the flow is potential upstream, we obtain the
following expression for the vorticity just behind the flame front
\begin{eqnarray}\label{vort3}&&
\sigma_{+} = - \frac{\theta - 1}{2\theta N}(u_{-}^2 + w_{-}^2)'.
\end{eqnarray}
\noindent With the help of this equation, we can now show that
Eq.~(\ref{flow3}) is actually satisfied by the model solution with
the accuracy of the sixth order in $\alpha \equiv \theta - 1,$ as
$\alpha \to 0.$

First of all, the following estimates can be readily obtained
\cite{siv2}:
$$f' =  O(\alpha),
~~~\tilde{u}, w \sim (f')^2= O(\alpha^2),$$ and, more generally,
$$\frac{d F}{d\eta} = O(\alpha) O(F),$$ for any functional $F =
F[f(\eta)],$ since the amplitude $A$ of a perturbation of the
flame front with the wavelength $\lambda$ is of the order
$\alpha\lambda$ \cite{zel1,zel2,siv2}. In particular, it follows
from Eq.~(\ref{vort3}) that\footnote{In the general 3D case, the
same result follows from the formula
$$\{{\rm rot}{\bf v}\}_{-}^{+}
= \frac{(\theta - 1)v^{t}_{-}}{\theta N}\left[{\bf n},
\nabla_t{\bf v}_{-}^{t}\right],$$ where $[~,~]$ denotes the vector
product, ${\bf n}$ the unit vector normal to the flame front
(pointing to the burnt matter), ${\bf v}^{t}_{-}$ the tangential
to the flame front component of the velocity, and $\nabla_t$
differentiation in the direction ${\bf v}_{-}^{t}$.}
\begin{eqnarray}\label{vort4}&&
\sigma = O(\alpha^4).
\end{eqnarray}
\noindent

Next, we rewrite Eq.~(\ref{flow3}) identically
\begin{eqnarray}\label{flow3n}&&
u \frac{\partial u}{\partial\eta} + w \frac{\partial
w}{\partial\eta} - u\sigma = - \frac{1}{\rho}\frac{\partial
\Pi}{\partial\eta}.
\end{eqnarray}
Setting $\xi = f(\eta)+ 0$ in this equation and using
Eq.~(\ref{vort3}), we obtain
\begin{eqnarray}\label{flow3n1}&&
\left(\frac{\partial}{\partial\eta}\frac{u^2 + w^2}{2}\right)_{+}
+ u_{+}\frac{\theta - 1}{2\theta N}\frac{d}{d\eta}(u_{-}^2 +
w_{-}^2) = - \theta\left(\frac{\partial
\Pi}{\partial\eta}\right)_{+}.
\end{eqnarray}
\noindent Using the above estimates and taking into account that
$u_{+} = \theta V + \tilde{u}_{+},$ $V = 1 + O(\alpha^2),$
Eq.~(\ref{flow3n1}) can be rewritten as
\begin{eqnarray}\label{flow3n2}&&
\left(\frac{\partial}{\partial\eta}\frac{u^2 + w^2}{2}\right)_{+}
+ (\theta - 1)\frac{d}{d\eta}\frac{u_{-}^2 + w_{-}^2}{2} = -
\theta\left(\frac{\partial\Pi}{\partial\eta}\right)_{+} +
O(\alpha^6).
\end{eqnarray}
\noindent On the other hand, similar transformations of
Eq.~(\ref{flow2}) give, with the same accuracy,
\begin{eqnarray}\label{flow2n}&&
\left(\frac{\partial}{\partial\xi}\frac{u^2 + w^2}{2}\right)_{+} =
- \theta\left(\frac{\partial \Pi}{\partial\xi}\right)_{+} +
O(\alpha^6).
\end{eqnarray}
\noindent Finally, taking the sum of Eq.~(\ref{flow3n2}) and
Eq.~(\ref{flow2n}) multiplied by $f',$ and noting that
$$u_{+}^2 + w_{+}^2 = u_{-}^2 + w_{-}^2 + \theta^2 - 1, ~~\Pi_{+} = \Pi_{-} - (\theta - 1),$$
we get
\begin{eqnarray}\label{flow3n3}&&
\frac{d}{d\eta}\left\{\frac{1}{2}\left(u_{-}^2 + w_{-}^2\right) +
\Pi_{-} \right\} = O(\alpha^6).
\end{eqnarray}
\noindent Since the flow is potential upstream, the left hand side
of  Eq.~(\ref{flow3n3}) is zero. We conclude that
Eq.~(\ref{flow3n3}), and therefore Eq.~(\ref{flow3}), is satisfied
with the accuracy of the sixth order in $\alpha.$

It is worth to emphasize that the model equation (\ref{model}) has
not been used in the derivation of Eq.~(\ref{flow3n3}). Therefore,
the latter holds true whatever model is considered, provided that
this model respects all the conservation laws at the flame front,
as well as the flow equations up- and downstream.  Furthermore,
Eq.~(\ref{model}) is model-independent up to the second order in
$\alpha,$ since to this order the flow is potential downstream.
Extended to all values of $\theta,$ Eq.~(\ref{model}) thus
provides the simplest model satisfying the above-mentioned
requirements.

Finally, considered as an equation for the quantity $(u_{-}^2 +
w_{-}^2)',$ Eq.~(\ref{flow3n3}) determines it with the accuracy of
$\alpha^6.$ On the other hand, since the left hand side of
Eq.~(\ref{main}) is proportional to the same quantity, $\alpha^6$
is the accuracy estimate for Eq.~(\ref{main}) as well.

It remains only to make the following three important remarks.

R1) The above accuracy estimate is obtained from the analysis of
differential equations governing the fluid dynamics. These
equations should be complemented by appropriate boundary
conditions. As we have already mentioned, the problem of flame
propagation is essentially nonlocal; this non-locality shows
itself in the fact that the boundary conditions for the burnt
matter, together with the boundary conditions for the flame front
itself, are invoked in the course of integration of
Eq.~(\ref{main}). By itself, this equation is independent of the
boundary conditions, since it is obtained by the direct
substitution of Eq.~(\ref{consist7}) into the Euler equation
(\ref{flow2}) for the burnt matter, written in the form
(\ref{phase2}). Thus, the fact that the consistency condition
(\ref{consist}) can be resolved with respect to the phase $\phi$
is crucial for the above accuracy estimate. How the boundary
conditions are actually taken into account in the course of
integration of Eq.~(\ref{main}) will be shown in
Sec.~\ref{4order}. Closely connected to this is the remark

R2) It was mentioned in Sec.~\ref{deriv} that the value
(\ref{constant}) of the constant $C$ entering Eq.~(\ref{ampl4}) is
fixed up to the second order, since to this order the flow is
potential downstream. One might think that the ambiguity in $C$ at
higher orders spoils the above $O(\alpha^6)$ accuracy estimate for
Eq.~(\ref{main}). It is not difficult to see, however, that within
its accuracy, Eq.~(\ref{main}) is not affected by this ambiguity.
Indeed, since the numerator in its right hand side is of the order
$\alpha^3,$ a third order change in $C$ in the denominator gives
rise only to terms of the order $\alpha^6,$ and the same is true
for the left hand side. This is what we should have expected,
since as we proved above, Eq.~(\ref{main}) is the true equation
with the accuracy $\alpha^6,$ and as it is it must be independent
of the model particularities in the higher-order completion of
$C.$

R3) As we mentioned in the points R1), R2), equation (\ref{main})
depends neither on the boundary conditions, nor on the higher
order completion of the constant $C.$ However, given the boundary
conditions, $C$ is fixed upon integration of Eq.~(\ref{main}). The
point is that the boundary conditions for the flame front together
with the boundary condition for the fluid velocity downstream
imply more strong restriction on the value of $C$ than that given
by Eq.~(\ref{constant}) to the second order in $\alpha.$ This is
because the flow structure near the ending points of the flame
front at the tube walls can be completely determined with the help
of the boundary conditions. Thus, the initial choice of the
constant $C$ is effectively corrected in the course of integration
of Eq.~(\ref{main}) by the choice of the integration constants,
appropriate to the given boundary conditions. In view of this
fact, in practice, it is more convenient to work with
Eq.~(\ref{consist7}), the undifferentiated version of
Eq.~(\ref{main}), from the very beginning, and choose the constant
$C$ from the requirement that the amplitude $\Omega$ and phase
$\phi$ take the boundary values prescribed by the boundary
conditions. We follow this way in Sec.~\ref{perexpansion} below.
Yet another example of determining the constant $C$ is given in
Sec.~\ref{largev}.

\subsection{Account of the finite flame thickness}\label{account}

We will show in this section how the considerations of Sec.~\ref{deriv}
can be generalized to take into account the effects due to finite
flame thickness.

In the case of small but nonzero flame thickness, the conservation
laws at the flame front read \cite{matalon}
\begin{eqnarray}\label{conserv1nn}&&
u_{+} - u_{-} - f'(w_{+} - w_{-}) = (\theta - 1) N,
\end{eqnarray}
\begin{eqnarray}\label{conserv2nn}&&
w_{+} - w_{-} + f'(u_{+} - u_{-}) =
\varepsilon\ln\theta\left(\hat{D}w_{-} + f'\hat{D}u_{-} +
\frac{1}{N}\hat{D}f' \right),
\end{eqnarray}
\begin{eqnarray}\label{conserv3nn}&&
\Pi_{+} - \Pi_{-} = - (\theta - 1) + \varepsilon (\theta -
1)\left(\frac{f'}{N}\right)' \nonumber\\&& +
\frac{\varepsilon\ln\theta}{N}\left(w_{-}^2 f'' + 2 \hat{D}N -
\frac{f' N'}{N}\right),
\end{eqnarray}
\noindent
while the evolution equation
\begin{eqnarray}\label{evolutioneqq}&&
u_{-} - f'w_{-} = N - \varepsilon\frac{\theta\ln\theta}{\theta -
1} \frac{d}{d\eta}\left(N w_{-} + f'\right),
\end{eqnarray}
\noindent where $\varepsilon$ is the small dimensionless ratio of
the flame thickness to the characteristic length of the problem,
and operator $\hat{D}$ is defined in Eq.~(\ref{operator}).

As in Sec.~\ref{deriv}, one should first exclude $f(\eta)$ from
the arguments of the flow variables entering these equations.
Using the continuity equation (\ref{flow1}) and taking into
account potentiality of the flow upstream, we write
\begin{eqnarray}&&
\frac{d w_{-}}{d\eta} = \left(\frac{\partial
w}{\partial\eta}\right)_{-} + \left(\frac{\partial w
}{\partial\xi}\right)_{-}\cdot f' = \left(\frac{\partial
w}{\partial\eta}\right)_{-} + \left(\frac{\partial u
}{\partial\eta}\right)_{-}\cdot f', \label{auxil1}\\&&
 \hat{D}w_{-} = \left(\hat{D}w\right)_{-} +
\left(\frac{\partial w }{\partial\xi}\right)_{-}\cdot\hat{D}f =
\left(\hat{D}w\right)_{-} + \left(\frac{\partial u
}{\partial\eta}\right)_{-}\cdot\hat{D}f, \label{auxil2}\\&&
\hat{D}u_{-} = \left(\hat{D}u\right)_{-} + \left(\frac{\partial u
}{\partial\xi}\right)_{-}\cdot\hat{D}f = \left(\hat{D}u\right)_{-}
- \left(\frac{\partial w }{\partial\eta}\right)_{-}\cdot\hat{D}f
\label{auxil3}
\end{eqnarray}
\noindent Introducing notation for the flow variables as in
Sec.~\ref{deriv}, the above equations can be written in the form
of the set ($*$), namely, up to the terms of the fourth order
(which turns out be sufficient for expanding Eq.~(\ref{main}) up
to the fifth order),
$$\left(
\begin{array}{rcl}
u^{+} - u^{-}&=& (\theta - 1)/N\\
w^{+} - w^{-}&=& - f'(\theta - 1)/N + \varepsilon\ln\theta f'f''\\
\Pi^{+} - \Pi^{-}&=& - (\theta - 1) + \varepsilon (\theta - 1) f''\\
u^{-} - f'w^{-}&=& N - \varepsilon \theta\ln\theta/(\theta - 1)
(\partial w^{-}/\partial\eta + \partial u^{-}/\partial\eta f' + f'')\\
\end{array}
\right)_{\xi = f(\eta),} \eqno (**)$$ all other equations
remaining the same as in the set ($*$). Following the reasoning of
Sec.~\ref{deriv}, all the flow variables can now be considered
$\xi$-independent, which fact is expressed by the special
designation $(\upsilon, \omega, \pi)$ for the variables $(u, w ,
\Pi),$ respectively.

Thus, we see that in the case of nonzero flame thickness,
Eq.~(\ref{ampl4}) for the amplitude of the potential mode modifies
to
\begin{eqnarray}\label{ampl4ep}
\Omega^2 = (\upsilon_p^{+})^2 + (\omega_p^{+})^2 =
\theta\{(\upsilon^{-})^2 + (\omega^{-})^2\} - 2\varepsilon\theta
(\theta - 1)f'' + \theta(\theta - 1).
\end{eqnarray}
\noindent The corresponding expression for the rate of change of
the phase $\phi,$ which we do not write explicitly because of its
complexity, can be obtained by substituting the equations ($**$)
into Eq.~(\ref{phase1}). Differentiating Eq.~(\ref{consist7}), and
using the expression for $\phi'$ together with Eq.~(\ref{ampl4ep})
for the amplitude, one obtains an equation for the function
$f(\eta),$ which generalizes Eq.~(\ref{main}) for the case of
nonzero flame thickness.

The accuracy analysis for this equation is very complicated and
will not be carried out here. It should be noted, however, that
such an analysis is superfluous to a considerable extent. Indeed,
the finite flame thickness is mainly taken into account in order
to provide a short wavelength cutoff for the spectrum of the flame
front perturbations, which insures the existence of a stationary
flame configuration. On the other hand, to the leading order in
$\alpha,$ the form of the $\varepsilon$-corrections to the
equation for $f(\eta)$ is known in advance from the linear theory
of Pelce and Clavin \cite{pelce}. It will be shown in the next
section that, along with the nonlinear $\varepsilon$-corrections,
the equations obtained above correctly reproduce their result for
the cutoff wavelength.

\section{The small $(\theta - 1)$ expansion}\label{perexpansion}

In its general form, Eq.~(\ref{main}) is very complicated. It is a
highly nonlinear integro-differential equation, which can be
solved only numerically. On the other hand, as we showed in
Sec.~\ref{estimate}, this equation correctly approximates the
exact equation for the flame front of zero thickness up to the
fifth order in $\alpha.$ We now turn to expanding
Eq.~(\ref{main}), generalized to the case of nonzero flame
thickness, up to this order. As the result, a much simpler
equation will be obtained, which generalizes the Sivashinsky
equation \cite{siv2} taking into account vorticity production in
the flame.

\subsection{Fourth order equation for the flame front}\label{4order}

We will see below that to the order being considered,
Eq.~(\ref{phase1}) can be integrated with respect to the phase,
which implies that both sides of Eq.~(\ref{main}) are actually
full derivatives. However, it would be hasty to simply omit these
derivatives: integration of Eq.~(\ref{main}) requires careful
account of the boundary conditions. Since this equation is
integral, it is not sufficient to impose only one condition to fix
the constants of integration in its left and right hand sides.
Boundary conditions for the burnt matter and for the flame front
itself actually supply two independent conditions to be used to
fix two arbitrary constants: one additive constant in the phase of
the complex function (\ref{complex}), and one multiplicative
constant in its amplitude.

We proved in Sec.~\ref{estimate} that within the accuracy of the
sixth order in $\alpha,$ Eq.~(\ref{main}) is independent of a
particular completion of the constant (\ref{constant}) beyond the
second order. However, different choices of $C$ correspond to
different values of the integration constants in the phase and
amplitude. It turns out that in the case of ideal tube walls, the
particular choice (\ref{constant}) made in Eq.~(\ref{main})
implies that integration of this equation gives exactly
Eq.~(\ref{consist7}) where $\phi_0 = 0,$ $\phi_0$ being the value
of the phase $\phi$ at the tube walls. We will prove this
statement only for zero-thickness flames. As we pointed out in
Sec.~\ref{account}, nonzero flame thickness should be taken into
account mainly in order to provide the short wavelength cutoff for
the flame front perturbations. On the other hand,
$\varepsilon$-dependence of the constants of integration is of
little interest, and will be neglected in what follows. Let us now
turn to the proof of the above statement. First of all, since the
downstream flow is potential up to the second order, ${\bf v}^+ =
{\bf v}_p^+ + O(\alpha^3),$ the phase $\phi_0 = {\rm arg}(u_p^+ +
i w_p^+) = O(\alpha^3),$ since $w^+ = 0$ at the walls.
Furthermore, noting that $u_{-}^2 + w_{-}^2 = ({\bf v_{-}\cdot
t})^2 + ({\bf v_{-}\cdot n})^2,$ where ${\bf t}$ and ${\bf n}$ are
the tangential and the normal to the flame front unit vectors,
respectively, taking into account that ${\bf v_{-}\cdot t}N =
w_{-} + f' u_{-},$ ${\bf v_{-}\cdot n} = 1,$ and using
Eq.~(\ref{vort3}) we obtain
$$\sigma = - \frac{\theta - 1}{2\theta N}\frac{d}{d\eta}
\frac{(w_{-} + f' u_{-})^2}{N^2}.$$ It follows from this formula
that $\sigma = 0$ at the walls, since the boundary conditions are
assumed ideal. We conclude that the flow is potential near the
walls, $u^+ = u_p^+,$ $w^+ = w_p^+,$ and thus $\phi_0 = 0.$ For
the same reason, the amplitude $\Omega = \theta$ at the walls. Now
recall that the value (\ref{constant}) for the constant $C$ was
obtained in Sec.~\ref{deriv} from the boundary conditions $u_p^{-}
= 1,$ $u_p^{+} = \theta,$ $w_p^{-} = w_p^{+} = 0,$ valid up to the
second order. We see, therefore, that in the case when the
boundary condition for the flame front is $f'=0,$
Eq.~(\ref{constant}) is extended to all orders in $\alpha,$ and
thus Eq.~(\ref{consist7}) is indeed the integral of
Eq.~(\ref{main}), satisfying the boundary conditions.

Let us now proceed with expanding Eq.~(\ref{consist7}) in powers
of $\alpha.$ Since the numerator in the right hand side of
Eq.~(\ref{phase1}) is of the third order in $\alpha,$ it is
sufficient to expand the denominator up to the second order:
\begin{eqnarray}&&
\left\{(V + \upsilon)^2 + (\hat{H}\upsilon)^2 - 2\varepsilon
(\theta - 1)f'' + (\theta - 1)\right\}^{- 1} \nonumber\\&& = 1 - 2
W - 2\upsilon - (\theta - 1) + (\theta - 1)^2 + O(\alpha^3),
\end{eqnarray}
where $W\equiv V - 1 = O(\alpha^2)$ is the flame velocity increase
due to the front curvature, and the designation
$\tilde{\upsilon}^{-}$ is again reduced to $\upsilon,$ for
brevity. Substituting this into Eq.~(\ref{phase1}), we find
\begin{eqnarray}\label{phase3}&& \frac{d\phi}{d\eta} =
\frac{1}{\theta}(1 - W -
\upsilon)\frac{d}{d\eta}\left(\hat{H}\upsilon - f'\frac{\theta -
1}{N} + \varepsilon\ln\theta f' f'' \right) \nonumber\\&& -
\frac{1}{\theta}\left\{\hat{H}\upsilon - f'(\theta -
1)\right\}\frac{d\upsilon}{d\eta} + O(\alpha^6).
\end{eqnarray}

It is not difficult to see that Eq.~(\ref{phase3}) can be
integrated to give
\begin{eqnarray}\label{phase4}&&
\phi = - \frac{1}{\theta}(W + \upsilon )\left(\hat{H}\upsilon -
f'(\theta - 1)\right) \nonumber\\&& +
\frac{\hat{H}\upsilon}{\theta} - f'\frac{\theta - 1}{\theta N} +
\frac{\varepsilon\ln\theta}{\theta} f' f'' + \phi_0 + O(\alpha^5),
\end{eqnarray}
\noindent where $\phi_0 = 0$ for ideal tube walls, as we have
shown above. One should remember, however, that $\phi_0$ assumes
different values under different boundary conditions. Thus,
$\phi_0 = O(\varepsilon\alpha^3),$ in general. It is worth to note
also that this dependence of the flame equation on the boundary
conditions for the burnt matter is a reflection of the essential
non-locality of the process of curved flame propagation, mentioned
already in the Introduction and connected to the vorticity drift
behind the flame front.

Next, we expand the right hand side of Eq.~(\ref{consist7})
\begin{eqnarray}\label{ampl5}&& - \hat{H}\ln\frac{\Omega}{\theta
V} = - \hat{H}\left\{ \frac{1}{2}\ln\left[1 + \frac{2}{\theta}(W +
\upsilon) + \frac{(W + \upsilon)^2}{\theta} +
\frac{(\hat{H}\upsilon)^2}{\theta} \right.\right.\nonumber\\&&
\left.\left. - 2\varepsilon\frac{(\theta - 1)}{\theta}f'' \right]
+ \ln\frac{1}{1 + W} \right\} = - \hat{H}\left\{\frac{(W +
\upsilon)}{\theta} + \frac{(\hat{H}\upsilon)^2 - (W + \upsilon
)^2}{2} \right.\nonumber\\&&\left.- \varepsilon \frac{(\theta -
1)}{\theta} f'' - W + \frac{W^2}{2} \right\} + O(\alpha^5).
\end{eqnarray}
\noindent

Substituting Eqs.~(\ref{phase4}), (\ref{ampl5}) into
Eq.~(\ref{consist7}) and rearranging with the help of the identity
(\ref{h8}) gives
\begin{eqnarray}\label{consist8}&&
\upsilon = \frac{\theta - 1}{2}\left(W -
\hat{H}\frac{f'}{N}\right) + \frac{\varepsilon}{2}\{(\theta - 1)
f'' + \ln\theta \hat{H}(f' f'')\} + O(\alpha^5).
\end{eqnarray}
\noindent On the other hand, one has from the last equation of the
set ($**$)\footnote{From now on, the terms $O(\alpha^5)$ will be
omitted, for simplicity.}
\begin{eqnarray}\label{evolutioneq}&&
(1 - f'\hat{H})\upsilon = N - V - \varepsilon
\frac{\theta\ln\theta}{\theta - 1} (\hat{H}\upsilon' + \upsilon'f'
+ f''),
\end{eqnarray}
\noindent which is the generalization of Eq.~(\ref{expansion1}) to
the case of nonzero flame thickness.

In the lowest, second, order in $\theta - 1,$
Eqs.~(\ref{consist8}) and (\ref{evolutioneq}) give, respectively,
\begin{eqnarray}\label{lowest1}&&
\upsilon = - \frac{\theta - 1}{2}\hat{H}f',
\end{eqnarray}
\begin{eqnarray}\label{lowest2}&&
\upsilon = \frac{(f')^2}{2} - W - \varepsilon f''.
\end{eqnarray}
The following equation for the flame front position, implied by
Eqs.~(\ref{lowest1}), (\ref{lowest2}),
\begin{eqnarray}\label{lowest}&&
\frac{(f')^2}{2} - W = - \frac{\theta - 1}{2}\hat{H}f' +
\varepsilon f''
\end{eqnarray}
\noindent is nothing but the stationary part of the Sivashinsky
equation \cite{siv2}.

In the next order, Eq.~(\ref{consist8}) becomes
\begin{eqnarray}\label{lowest3}&&
\upsilon = \frac{\theta - 1}{2}(W - \hat{H}f') +
\frac{\varepsilon}{2}f''.
\end{eqnarray}
\noindent Substituting this into Eq.~(\ref{evolutioneq}) and
expanding to the third order, we obtain the following equation
\begin{eqnarray}\label{lowestnew}&&
\frac{\theta}{2}(f')^2 - \frac{\theta + 1}{2}W = \frac{\theta -
1}{2}\left( - \hat{H}f' + \frac{\lambda^{(1)}}{2\pi} f''\right),
\\&& \lambda^{(1)}\equiv \frac{4\pi\varepsilon}{\theta -
1}\left(1 + \frac{3(\theta - 1 )}{2}\right).\label{lambda1}
\end{eqnarray}
\noindent Note that the expression (\ref{lambda1}) is just the
first order approximation to the exact value of the cutoff
wavelength
\begin{eqnarray}
\lambda \equiv \frac{2\pi\varepsilon}{\theta - 1
}\left(\theta\ln\theta\frac{\theta + 1}{\theta - 1 } + \theta -
1\right),
\end{eqnarray}
\noindent given by the linear theory of the LD-instability
\cite{pelce}.

Finally, at the fourth order, substituting Eq.~(\ref{consist8})
into Eq.~(\ref{evolutioneq}), and using the lower order equations
(\ref{lowest}) and (\ref{lowestnew}), we obtain
\begin{eqnarray}\label{lowestnew4}&&
\hspace{-1cm}\frac{\theta}{2}(f')^2 - \frac{\theta + 1}{2}W -
\frac{(f')^4}{8} + \frac{\theta - 1}{4}\left\{f'\hat{H}(f')^2 -
\hat{H}(f')^3\right\} - \frac{(\theta - 1)^2}{4}(f')^2
\nonumber\\&& \hspace{-1cm}= - \frac{\theta - 1}{2}\hat{H}f' +
\frac{\varepsilon}{2}\left(\theta\ln\theta\frac{\theta + 1}{\theta
- 1 } + \theta - 1\right)f'' + \varepsilon\frac{\theta -
1}{2}\left\{\hat{H}(f' f'') - f'\hat{H}f''\right\}.
\end{eqnarray}
\noindent We prefer not to expand the logarithm in the right hand
side of this equation to make it transparent that the linear terms
in Eq.~(\ref{lowestnew4}) again correctly reproduce the
corresponding terms of the linear theory.

\subsection{Solution of the third order equation for the flame
front}\label{3order}

The third order equation (\ref{lowestnew}) is of the same
functional structure as the Sivashinsky equation (\ref{lowest}).
Therefore, it can be solved analytically using the method of pole
decomposition \cite{henon,joulin}. Considering the flame
propagation in a tube of width $b$ with ideal walls (for
definiteness, the walls are taken to be the lines $\eta = 0$ and
$\eta = b$), we look for $2 b$-periodic solutions of the form
\begin{eqnarray}\label{anzats} f(\eta) = a \sum_{k = 1}^{2 P}
\ln\sin\left[\frac{\pi}{2 b}(\eta - \eta_k)\right],
\end{eqnarray}
\noindent where the amplitude $a$ and the complex poles $\eta_k,$
$k = 1,...,2P$ are to be determined substituting this anzats into
Eq.~(\ref{lowestnew}). Since the function $f(\eta)$ is real for
real $\eta,$ the poles come in conjugate pairs; $P$ is the number
of the pole pairs. Requiring the $2 b$-periodic solutions to be
symmetric with respect to the reflection $\eta \to - \eta,$ one
can obtain periodic as well as non-periodic solutions to
Eq.~(\ref{lowestnew}) in the domain $\eta \in (0,~b)$, satisfying
the conditions $f'(\eta = 0) = f'(\eta = b) = 0.$

Using the formulae\footnote{Since the application of pole
decomposition to Eq.~(\ref{lowestnew}) is quite similar to that
given in Refs.~\cite{henon,joulin}, we refer the reader to these
works for more detail.}
\begin{eqnarray}&&\label{fjoul}
\hat{H}f' = - \frac{\pi a}{2 b}\sum_{k = 1}^{2 P}\left\{1 + i~{\rm
 sign}({\rm Im}~\eta_k)\cot\left[\frac{\pi}{2 b}(\eta -
\eta_k)\right]\right\}, ~{\rm sign}(x) \equiv
\frac{x}{|x|},\nonumber\\&& \cot x \cot y = -1 + \cot(x - y )(\cot
y - \cot x),
\end{eqnarray}
\noindent it is not difficult to verify that Eq.~(\ref{lowestnew})
is satisfied by $f(\eta)$ taken in the form of Eq.~(\ref{anzats}),
provided that
\begin{eqnarray}&&\label{solution1}
a = - \frac{2\varepsilon}{\theta}\left(1 + \frac{3(\theta - 1
)}{2}\right), \nonumber\\&& W = \frac{(\theta - 1)^2}{\theta
(\theta + 1)}\frac{P\lambda^{(1)}}{2 b}\left(1 -
\frac{P\lambda^{(1)}}{2 b}\right),
\end{eqnarray}
\noindent and the poles $\eta_k,$ $k = 1,...,2P,$ satisfy the
following set of equations
\begin{eqnarray}&&\label{solution2}
i~{\rm sign}({\rm Im}~\eta_k) + \frac{\lambda^{(1)}}{2
b}\sum\limits_{m = 1 \atop m\ne k}^{2 P}\cot\left[\frac{\pi}{2
b}(\eta_k - \eta_m)\right] = 0, ~k = 1,...,2P.
\end{eqnarray}
\noindent It is seen from Eq.~(\ref{solution1}) that for the tube
width $b>\lambda^{(1)},$ the solution (\ref{anzats}) is not
unique: different solutions corresponding to different numbers $P$
of poles are possible. To find the physical ones, the stability
analysis is required which, of course, cannot be carried out in
the framework of the stationary theory. However, as we have
mentioned above, the functional structure of Eq.~(\ref{lowestnew})
is very similar to that of the stationary Sivashinsky equation
(\ref{lowest}). Under assumption that the non-stationary versions
of these equations are also similar, the stability analysis of
Refs.~\cite{siv3}--\cite{siv5} will be carried over the present
case. According to this analysis, for a given tube of sufficiently
small width, there is only one (neutrally) stable solution. This
solution corresponds to the number of poles that provides maximal
flame velocity. In addition to that, the pole structure of the
stable solution is such that the poles form a vertical alignment
in the complex $\eta$-plane, sharing the the same common real
part. For such a "coalescent" solution, a simple upper bound on
the number of poles can be obtained from Eq.~(\ref{solution2}).
Namely, for $k = k_0$ with $\eta_{k_0}$ uppermost, one has
\begin{eqnarray}&&\label{solution3}
1 = \frac{\lambda^{(1)}}{2 b}\sum\limits_{m = 1 \atop m\ne
k_{0}}^{2 P}\coth\left[\frac{\pi}{2 b}({\rm Im}~\eta_{k_0} - {\rm
Im }~\eta_m)\right] \ge \frac{\lambda^{(1)}}{2 b}(2 P - 1).
\nonumber
\end{eqnarray}
\noindent (The equality holds, if ${\rm Im}~\eta_{k_0} = \infty$)
Then it follows from Eq.~(\ref{solution1}) that the maximum of the
flame velocity corresponds to the maximal number of the pole pairs
$$P_{{\rm max}} = {\rm Int} \left(\frac{b}{\lambda^{(1)}} + \frac{1}{2}\right),$$
${\rm Int}(x)$ denoting the integer part of $x.$ Thus, the flame
velocity increase $W_s$ of the stable solution can be written as
\begin{eqnarray}&&\label{solution4}
W_s = 4 W_{{\rm max}}\frac{P_{{\rm max}}\lambda^{(1)}}{2 b}\left(1
- \frac{P_{{\rm max}}\lambda^{(1)}}{2 b}\right),
\end{eqnarray}
\noindent where
\begin{eqnarray}\label{wmax}
W_{{\rm max}} = \frac{(\theta - 1)^2}{4\theta
(\theta + 1)}.
\end{eqnarray}

\begin{figure}
\hspace{3cm} \epsfxsize=8,5cm\epsfbox{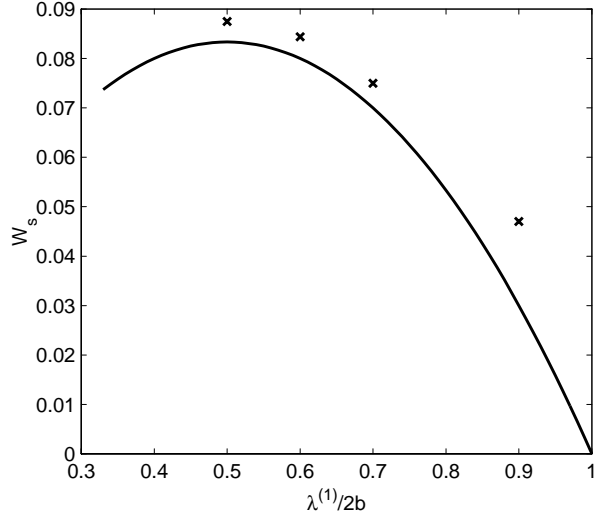}
\caption{Dependence of the flame velocity increase on the inverse
tube width, given by Eq.~(\ref{solution4}) for the case $\theta =
3$, and the results of numerical experiments [5].}\label{fig1}
\end{figure}

\begin{figure}
\hspace{3cm} \epsfxsize=8,5cm\epsfbox{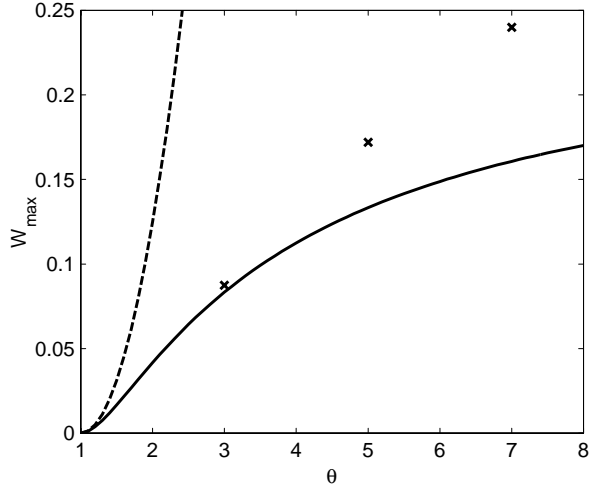}
\caption{Dependence of the maximal flame velocity increase
$W_{{\rm max}}$ on the gas expansion coefficient, given by
Eq.~(\ref{wmax}) (solid line), and calculated on the basis of the
Sivashinsky equation (dashed line); the marks are according to the
results of [5].}\label{fig2}
\end{figure}

Fig.~\ref{fig1} compares the theoretically predicted dependence of
the flame velocity increase on the inverse tube width, given by
Eq.~(\ref{solution4}) for the case $\theta = 3$, with the results
of numerical experiments \cite{bychkov2}. Dependence of the
maximal flame velocity increase $W_{{\rm max}}$ on the gas
expansion coefficient, given by Eq.~(\ref{wmax}), is represented
in Fig.~\ref{fig2}. For comparison, we show also the corresponding
dependence calculated with the help of the Sivashinsky equation.
We see that the third order equation (\ref{lowestnew}) gives
reasonable description of flames with $\theta \sim 3,$ while for
larger values of the expansion coefficient it overestimates the
influence of vorticity, produced in the flame, on the flame front
curvature. In the latter case, therefore, the more accurate fourth
order equation (\ref{lowestnew4}) should be used instead of
Eq.~(\ref{lowestnew}). Detailed investigation of
Eq.~(\ref{lowestnew4}) will be given elsewhere.

\section{Large flame velocity expansion}\label{largev}

As we have seen, Eq.~(\ref{main}) considerably simplifies in the
case of small $\alpha.$ For sufficiently narrow tubes, it gives
results which turn out to be in a reasonable agreement with the
experiment already at the third order in $\alpha.$ Let us now
consider the opposite case of very wide tubes, i.e., tubes of
width large compared to the cutoff wavelength. As it will be shown
presently, under a certain burning regime, Eq.~(\ref{main}) can be
written in a much simpler form in this case too.

Widely employed in modern jet engines is the process of the
so-called fast flow burning (see, e.g., \cite{zel2}, Ch.6,~$\S$1).
This regime is characterized by a large stretch of the flame front
along the tube, since the flame velocity is proportional to the
flame front surface (in 2D case, to the front length). Indeed,
equating the fuel flow at $\xi = - \infty$ to that through the
flame front, using the evolution equation (\ref{evolutioneqq}),
and taking into account the boundary conditions $w = 0,$ $f' = 0,$
one has
\begin{eqnarray}&&
b V = \int\limits_{0}^{b}d l ~{\bf n}{\bf v}_{-} =
\int\limits_{0}^{b}d\eta (u_{-} - f' w_{-}) =
\int\limits_{0}^{b}d\eta \left\{N
-\varepsilon\frac{\theta\ln\theta}{\theta - 1}
\frac{d}{d\eta}\left(N w_{-} + f'\right)\right\} \nonumber\\&& =
\int\limits_{0}^{b}d\eta N -
\varepsilon\left.\frac{\theta\ln\theta}{\theta - 1}\left(N w_{-} +
f'\right)\right|_{0}^{b} = l, \nonumber
\end{eqnarray}
\noindent or
$$V = \frac{l}{b} \gg 1,$$
where $l$ is the length of the flame front, ${\bf n}$ the unit
vector normal to the flame front (pointing to the burnt matter),
and $b$ is the tube width.

Under these circumstances, it is natural to develop an expansion
of Eq.~(\ref{main}) in powers of the inverse flame velocity,
$1/V.$ This can be done as follows.

\begin{figure}
\epsfxsize=6,5cm \epsfbox[-400, 0, 195, 842]{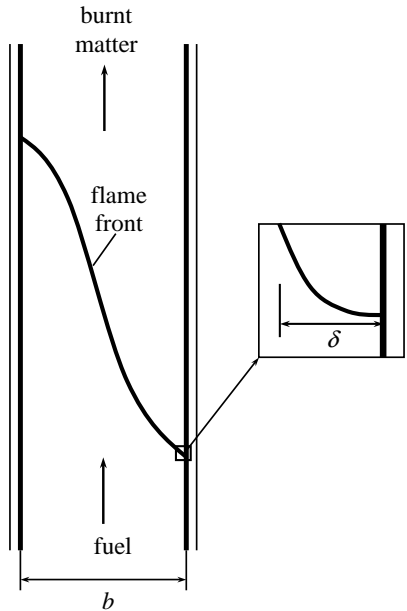}
\caption{Schematic representation of the flame front shape in the
fast flow burning regime.}\label{fig3}
\end{figure}

Let us note, first of all, that large value of $V$ implies, in
general, that the quantity $\upsilon$ (and, therefore, the
quantity $\hat{H}\upsilon$ -- the $\eta$-component of the fuel
velocity at the flame front) is also large. Indeed, as it follows
from the evolution equation (\ref{evolution}), $\upsilon = 1 - V$
at the tube wall, in view of the boundary condition $f'=0.$ Now,
we make the assumption that in the bulk, i.e., for all $\eta$
except for a small region near the points $\eta = 0,~b,$ the
quantity $\upsilon$ is actually of the order $O(1),$ rather than
$O(V).$ Then the integral quantity $\hat{H}\upsilon = O(1) +
\delta/b~O(V) = o(V),$ since the size $\delta$ of the region where
$\upsilon \sim V$ is assumed to be small compared with the tube
width $b,$ $\delta\ll b$ (see Fig.~\ref{fig3}).

Expansion of Eq.~(\ref{main}) in powers of $1/V$ is
straightforward. Below, this expansion will be carried out only to
the lowest nontrivial order. As in Sec.\ref{perexpansion},
Eq.~(\ref{phase}) for the phase integrates at this order.
Therefore, it is more convenient to work with the integral version
(\ref{consist7}) of the equation (\ref{main}) from the very
beginning.

Let us first consider the question concerning the form of the
quantities $\Omega,$ $\phi$ entering this equation. As we saw in
Sec.~\ref{deriv}, the value (\ref{constant}) of the constant $C$
in the expression (\ref{ampl4}) for the amplitude $\Omega$ is
fixed up to the second order in $\alpha.$ Furthermore, it was
shown in Sec.~\ref{estimate}, that within the accuracy of our
model, the form of Eq.~(\ref{main}) is independent of a particular
completion of $C$ beyond the order $\alpha^2,$ and that the value
of $C,$ together with values of the arbitrary constants of
integration of Eq.~(\ref{main}), can be deduced from the boundary
conditions. In contrast, the boundary conditions cannot be used
directly in the framework of the $1/V$-expansion, since this
expansion breaks down near the tube walls. Instead, the value of
the constant $C$ as well as the constant $\phi_0$ of integration
of Eq.~(\ref{phase1}) will be determined from the requirement of
self-consistency of the limiting transition $V\to \infty$ in
Eq.~(\ref{consist7}).

Under assumption $\upsilon, ~\hat{H}\upsilon \sim o(V),$ it
follows from Eq.~(\ref{phase2}) that $\phi' \to 0$ in the bulk
when $V\to\infty;$ therefore, $\phi = \phi_0 = {\rm const}.$ On
the other hand, the amplitude $\Omega \to \sqrt{\theta V^2 + C}$
in the same limit. Thus, Eq.~(\ref{consist7}) gives in the lowest
order in $1/V:$
$$\phi_0 = - \hat{H}\ln\frac{\sqrt{\theta V^2 + C}}{\theta V}.$$
This equality is only consistent with the properties of the
Hilbert operator if\footnote{The operator (\ref{h}) is badly
defined for $f(\eta) = C = {\rm const},$ except for $C = 0.$
However, one can formally extend this definition to include
nonzero values of $C$ by setting $\hat{H}C = \beta C$ with some
number $\beta$ which must be one and the same for all $C,$ in
order to preserve the linearity of the Hilbert operator.
Furthermore, the properties (\ref{h1}) and (\ref{h8}) are
preserved only if $\beta = \pm i.$ Since the quantities
$\phi,~\Omega$ are real by definition, the choice (\ref{choice})
is unique.}
$$\ln\frac{\sqrt{\theta V^2 + C}}{\theta V} = 0,$$ and therefore,
\begin{eqnarray}\label{choice}
\phi_0 = 0, ~~C = \theta (\theta - 1)V^2.
\end{eqnarray}
\noindent Note that in the case of small $\alpha,$ the values
(\ref{choice}) for the constants $C,$ $\phi_0$ coincide with those
found in Secs.~\ref{deriv}, \ref{4order}, respectively, up to the
order $\alpha^2$. On the other hand, expansion of
Eq.~(\ref{consist7}) in powers of $\upsilon/V$ has the strict
validity in this case, since $\upsilon\sim O(\alpha^2),$ $V = 1 +
O(\alpha^2)$ for all $\eta \in (0,~b).$ We conclude, therefore,
that the first order of the large $V$ expansion of
Eq.~(\ref{consist7}) must automatically reproduce the second order
of the small $\alpha$ expansion of this equation, i.e., the
Sivashinsky equation (\ref{lowest}).

Substituting Eq.~(\ref{choice}) into Eq.~(\ref{consist7}) and
expanding to the first order in $\upsilon/V$ gives
$$\frac{V\left(\hat{H}\upsilon - f'\frac{\theta - 1}{N}\right)}{\theta^2 V^2}
= - \hat{H}\frac{\upsilon}{\theta V},$$ or
\begin{eqnarray}\label{upsilon}
\upsilon = - \frac{\theta - 1}{\theta +
1}\hat{H}\left(\frac{f'}{N}\right).
\end{eqnarray}
\noindent Since the flame is highly stretched in the fast flow
burning regime, the slope $f'$ of the flame front is large.
However, it follows from Eq.~(\ref{upsilon}) that the quantities
$\upsilon$ and $\hat{H}\upsilon$ remain of the order O(1), since
$f'/N < 1.$ Our initial assumption is thus confirmed.

Substituting Eq.~(\ref{upsilon}) into the evolution equation
(\ref{evolutionn}), we obtain
\begin{eqnarray}\label{largemain}
N - V = - \frac{\theta - 1}{\theta + 1} (f' +
\hat{H})\left(\frac{f'}{N}\right).
\end{eqnarray}
\noindent We see that in the case of small $\theta - 1,$
Eq.~(\ref{largemain}) does reproduce the stationary part of the
Sivashinsky equation (\ref{lowest}) in the case $\varepsilon = 0.$

Finally, it is not difficult to take into account the influence of
the small flame thickness. First of all, we note from
Eq.~(\ref{upsilon}) that although the quantity $\upsilon = O(1),$
its $\eta$-derivative
$$\upsilon' = - \frac{\theta - 1}{\theta +
1}\hat{H}\left(\frac{f'}{N}\right)' = - \frac{\theta - 1}{\theta +
1}\hat{H}\left(\frac{f''}{N^3}\right) =
O\left(\frac{1}{V^2}\right),$$ since $f'', N\sim f'\sim V.$ It
follows then from Eqs.~(\ref{auxil1})--(\ref{auxil3}) that
$\hat{D}u_{-},~\hat{D}w_{-} = O(1/V).$ Resolving
Eqs.~(\ref{conserv1nn}), (\ref{conserv2nn}) with respect to
$u_+,~w_+,$ we see that the $\varepsilon$-corrections to these
quantities are only of the relative order $1/V^2.$ Analogously,
the $\varepsilon$-correction in the pressure jump
(\ref{conserv3nn}) is $O(1),$ which implies the same $O(1/V^2)$
relative correction in the amplitude $\Omega.$ Thus, we conclude
that Eq.~(\ref{consist7}) remains unchanged in the first order in
$1/V,$ and so does, therefore, its consequence,
Eq.~(\ref{upsilon}). Substituting the latter into evolution
equation (\ref{evolutioneqq}), we obtain
\begin{eqnarray}&&\label{largemainep} N - V = - \frac{\theta -
1}{\theta + 1} (f' + \hat{H})\left(\frac{f'}{N}\right) +
\varepsilon\frac{2\theta^2\ln\theta}{\theta^2 - 1} f''.
\end{eqnarray}
\noindent It is interesting to note that the term describing the
influence of finite flame thickness turns out to be linear in $f.$

\section{Discussion and conclusions}\label{conclud}

We have shown that in the stationary case, the asymptotic
expansion of the nonlinear equation for the flame front position
can be pushed beyond the second order in $\theta - 1,$ at which
the gas flow is potential on both sides of the flame front, to
take into account vorticity drift behind the flame front. This
expansion has been carried out explicitly; for the case of ideal
tube walls, it is given by Eqs.~(\ref{lowestnew}),
(\ref{lowestnew4}) at the third and fourth orders, respectively.
Remarkably, the third order equation, which describes influence of
the vorticity on the flame front structure to the lowest
nontrivial order, turns out to have the same functional structure
as the Sivashinsky equation. As we showed in Sec.~\ref{3order}, it
gives results in a reasonable agreement with the numerical
experiments on the flame propagation in tubes for the case of
flames with $\theta\le 3.$

It should be mentioned that the nonlinear equation for the flame
front, derived in Ref.~\cite{bychkov1}, also gives satisfactory
description of the stationary flame propagation in narrow tubes.
We showed already in the Introduction that the approach of
Ref.~\cite{bychkov1} is self-contradictory. Still, one might
imagine that despite an erroneous derivation, the resulting
equation is correct. However, the results of Sec.~\ref{4order}
show that in the case $\theta \to 1,$ asymptotic expansion of the
true nonlinear equation is quite different from that of the
equation proposed in Ref.~\cite{bychkov1}. The latter is
incorrect, therefore, already in the case of small gas expansion.

From a more general point of view, the theory of flame propagation
in the fully developed nonlinear regime cannot be formulated in
the way the Sivashinsky equation \cite{siv2} or the Frankel
equation \cite{frankel} are formulated. This is because the
assumption of potentiality of the flow downstream, employed in
these works, renders the relations between the flow variables
local, allowing thereby the formulation of equation for the flame
front position in terms of this position alone. In contrast,
account of the vorticity production in the flame makes the problem
essentially nonlocal, since, e.g., the value of the pressure field
on the flame front is a functional of the velocity field in the
bulk, which implies, in particular, that the equation for the
flame front position must be a functional of the boundary
conditions. Indeed, as we saw in Sec.~\ref{4order}, the boundary
conditions on the tube walls are invoked in the course of
derivation of the asymptotic expansion of this equation already at
the third order.

Finally, with the help of the general equation (\ref{main}),
highly nonlinear regimes of the stationary flame propagation can
be considered. We would like to remind that this equation respects
all the conservation laws to be satisfied across the flame front,
by construction. Thus, despite the fact that the vorticity drift
behind the flame front is taken into account in this equation on
the basis of the model assumption (\ref{model}), unjustified for
arbitrary $\theta,$ one may hope that it gives at least
qualitative description. We showed in Sec.~\ref{largev} that in
the particular case of the fast flow burning, equation
(\ref{main}) (as well as its generalization to the case of nonzero
flame thickness) can be highly simplified by expanding it in
powers of the inverse flame velocity. The result of this expansion
to the first order is given by Eq.~(\ref{largemainep}).

\begin{center}
{\large \bf Acknowledgements}
\end{center}

We are grateful to V.~V.~Bychkov for interesting discussions.

This research was supported in part by Swedish Ministry of
Industry (Energimyndigheten, contract P 12503-1), by the Swedish
Research Council (contract E5106-1494/2001), and by the Swedish
Royal Academy of Sciences.

\section{Appendix}

For the sake of completeness, we give here a brief account of the
properties of the Hilbert operator, used in the text.

Given a sufficiently smooth integrable function $f(\eta),$ $\eta
\in (-\infty,+\infty),$ the Hilbert operator $\hat{H}$ is defined
by
\begin{eqnarray}\label{h}
(\hat{H}f)(\eta) = \frac{1}{\pi}{\rm
p.v.}\int\limits_{-\infty}^{+\infty}d\zeta\frac{f(\zeta)}{\zeta -
\eta},
\end{eqnarray}
\noindent $"{\rm p.v.}"$ denoting the principal value. By
definition, operator $\hat{H}$ is linear, i.e., for any complex
numbers $c_1,$ $c_2,$
$$\hat{H}(c_1 f_1 + c_2 f_2) = c_1\hat{H}f_1 + c_2\hat{H}f_2,$$
and real, i.e.,
$$(\hat{H}f)^{*} = \hat{H}f^{*},$$ where $F^{*}$ denotes the
complex conjugate of $F.$

It is convenient to introduce the usual scalar product of two
functions $f_1(\eta)$ and $f_2(\eta)$
$$(f_1,f_2) = \int\limits_{-\infty}^{+\infty}d\eta f_1^{*}(\eta) f_2(\eta).$$
Then, changing the order of integration, one has
\begin{eqnarray}
(f_1,\hat{H}f_2) =
\frac{1}{\pi}\int\limits_{-\infty}^{+\infty}d\eta f_1^{*}(\eta)
{\rm
p.v.}\int\limits_{-\infty}^{+\infty}d\zeta\frac{f_2(\zeta)}{\zeta
- \eta} = - \frac{1}{\pi}\int\limits_{-\infty}^{+\infty}d\zeta
f_2(\zeta){\rm p.v.}
\int\limits_{-\infty}^{+\infty}d\eta\frac{f_1^{*}(\eta)}{\eta -
\zeta} = - (\hat{H}f_1,f_2), \nonumber
\end{eqnarray}
\noindent or
\begin{eqnarray}\label{herm}
(f_1,\hat{H}f_2) =
 - (\hat{H}f_1,f_2),
\end{eqnarray}
\noindent i.e., the Hilbert operator is anti-Hermitian.

\begin{figure}
\hspace{1,5cm} \epsfxsize=8,5cm\epsfbox[-20, 550, 300,
800]{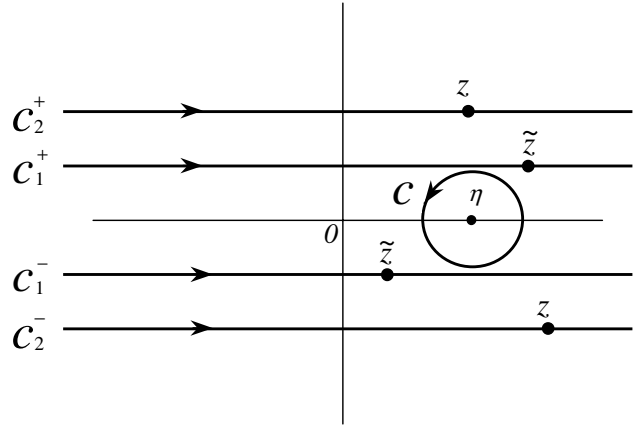} \caption{Contours of integration in
Eqs.~(\ref{h2})--(\ref{h6}).}\label{fig4}
\end{figure}

To prove the operator identity
\begin{eqnarray}\label{h1}
\hat{H}^2 = - 1,
\end{eqnarray}
\noindent it is convenient to represent the right hand side of
Eq.~(\ref{h}) as the integral over the contour $C_1 = C_1^- \cup
C_1^+$ in the complex $\eta$-plane (see Fig.~\ref{fig4})
\begin{eqnarray}\label{h2}
(\hat{H}f)(\eta) = \frac{1}{2\pi}\int\limits_{C_1}d z\frac{f(z)}{z
- \eta}.
\end{eqnarray}
\noindent Then the square of the Hilbert operator takes the form
\begin{eqnarray}\label{h3}
(\hat{H}^2 f)(\eta) = \frac{1}{4\pi^2}\int\limits_{C_1}\frac{d
\tilde{z}}{\tilde{z} - \eta}\int\limits_{C_2}d z\frac{f(z)}{z -
\tilde{z}},
\end{eqnarray}
\noindent where the contour $C_2 = C_2^- \cup C_2^+$ comprises
$C_1.$ Changing the order of integration in Eq.~(\ref{h3}), using
the formula
\begin{eqnarray}
\int\frac{d \tilde{z}}{(\tilde{z} - \eta)(z - \tilde{z})} =
\frac{1}{z - \eta}\ln\frac{\tilde{z} - \eta}{\tilde{z} - z},
\nonumber
\end{eqnarray}
\noindent and taking into account that the logarithm gives rise to
a nonzero contribution only if the arguments of the functions
$\tilde{z} - \eta$ and $\tilde{z} - z$ run in opposite directions
when $\tilde{z}$ runs the contours $C_1^{\pm},$ we obtain
\begin{eqnarray}&&
\frac{1}{4\pi^2}\int\limits_{C_1}\frac{d \tilde{z}}{\tilde{z} -
\eta}\int\limits_{C_2}d z\frac{f(z)}{z - \tilde{z}} =
\frac{1}{4\pi^2}\left(2\pi i\int\limits_{C_2^-}d z \frac{f(z)}{z -
\eta } - 2\pi i\int\limits_{C_2^+}d z \frac{f(z)}{z - \eta
}\right) \nonumber\\&& = \frac{i}{2\pi}\int\limits_C d z
\frac{f(z)}{z - \eta }, \nonumber
\end{eqnarray}
\noindent and therefore,
$$(\hat{H}^2 f)(\eta) = - f(\eta).$$

Next, let us consider the quantity
\begin{eqnarray}&&\label{h6}
(\hat{H}\{f\hat{H}f\})(\eta) =
\frac{1}{4\pi^2}\int\limits_{C_1}d\tilde{z}\frac{f(\tilde{z})}{\tilde{z}
- \eta}\int\limits_{C_2}d z\frac{f(z)}{z - \tilde{z}}
\nonumber\\&& =
\frac{1}{4\pi^2}\int\limits_{C_1}d\tilde{z}\frac{f(\tilde{z})}{\tilde{z}
- \eta}\int\limits_{C_2}d z\frac{f(z)}{z - \eta} -
\frac{1}{4\pi^2}\int\limits_{C_2}d z\frac{f(z)}{z -
\eta}\int\limits_{C_1}d\tilde{z}\frac{f(\tilde{z})}{\tilde{z} - z
}.
\end{eqnarray}
\noindent The first term in the right hand side of Eq.~(\ref{h6})
is just $\{(\hat{H}f)(\eta)\}^2.$ The second term can be
transformed as follows. The contour $C_1^-$ of the
$\tilde{z}$-integral can be moved down pass the contour $C_2^-,$
the pole at $\tilde{z} = z$ giving rise to the extra term $- 2\pi
i f(z),$ $z\in C_2^-.$ Likewise, $C_1^+$ can be moved up pass
$C_2^+,$ with the extra term $2\pi i f(z),$ $z\in C_2^+.$ Thus,
\begin{eqnarray}&&
\frac{1}{4\pi^2}\int\limits_{C_2}d z\frac{f(z)}{z -
\eta}\int\limits_{C_1}d\tilde{z}\frac{f(\tilde{z})}{\tilde{z} - z
} = \frac{1}{4\pi^2}\int\limits_{C_1}d z\frac{f(z)}{z -
\eta}\int\limits_{C_2}d\tilde{z}\frac{f(\tilde{z})}{\tilde{z} - z
} \nonumber\\&& - \frac{i}{2\pi}\left(\int\limits_{C_2^-}d
z\frac{f^2(z)}{z - \eta} - \int\limits_{C_2^+}d z\frac{f^2(z)}{z -
\eta}\right) =
\frac{1}{4\pi^2}\int\limits_{C_1}d\tilde{z}\frac{f(\tilde{z})}{\tilde{z}
- \eta}\int\limits_{C_2}d z\frac{f(z)}{z - \tilde{z}}
\nonumber\\&& - \frac{i}{2\pi}\int_C d z\frac{f^2(z)}{z - \eta} =
(\hat{H}\{f\hat{H}f\})(\eta) + f^2(\eta). \nonumber
\end{eqnarray}
\noindent Substituting this into Eq.~(\ref{h6}), we arrive at the
following identity
\begin{eqnarray}&&\label{h8}
2\hat{H}\{f\hat{H}f\} = (\hat{H}f)^2 - f^2.
\end{eqnarray}

Finally, let us establish the connection
\begin{eqnarray}&&\label{h9}
\hat{\Phi} = - \partial\hat{H}
\end{eqnarray}
between the 2D Landau-Darrieus operator $\hat{\Phi}$ and the
Hilbert operator. The former is defined by
\begin{eqnarray}&&\label{h10}
(\hat{\Phi}f)(\eta) = \int\limits_{- \infty}^{+ \infty} d k |k|
f_{k} e^{i k \eta},\nonumber
\end{eqnarray}
\noindent $f_k$ being the Fourier transform of $f.$ Using the
Fourier representation of the function $"{\rm sign}"$
$${\rm sign}(k) = \frac{1}{2\pi i}\int\limits_{-\infty}^{+\infty}e^{i k \zeta}
\left(\frac{1}{\zeta - i 0} + \frac{1}{\zeta + i 0}\right),$$ one
has
\begin{eqnarray}&&
(\hat{\Phi}f)(\eta) = \int\limits_{- \infty}^{+ \infty} d k~k~{\rm
sign}(k) f_{k}e^{i k \eta} \nonumber\\&& = -
\frac{1}{2\pi}\frac{\partial}{\partial\eta}\int\limits_{-
\infty}^{+ \infty} d\zeta \left(\frac{1}{\zeta - i 0} +
\frac{1}{\zeta + i
0}\right)\int\limits_{-\infty}^{+\infty}dk~f_{k} e^{i k (\eta +
\zeta)} \nonumber\\&& = -
\frac{1}{2\pi}\frac{\partial}{\partial\eta}\int\limits_{-\infty}^{+\infty}d\zeta
f(\eta + \zeta)\left(\frac{1}{\zeta - i 0} + \frac{1}{\zeta + i
0}\right) = - \frac{1}{\pi}\frac{\partial}{\partial\eta}{\rm
p.v.}\int\limits_{-\infty}^{+\infty}d\zeta\frac{f(\eta +
\zeta)}{\zeta},\nonumber
\end{eqnarray}
\noindent or
$$(\hat{\Phi}f)(\eta) = - \frac{\partial}{\partial\eta}(\hat{H}f)(\eta).$$

Eq.~(\ref{h9}) is convenient in estimating the orders of magnitude
of expressions involving the operator $\hat{\Phi}.$ In view of
Eqs.~(\ref{herm}) and (\ref{h1}), one has
$$(\hat{H}f,\hat{H}f) = - (\hat{H}^2 f,f) = (f,f),$$ i.e., the Hilbert operator
has unit norm, and therefore
$$\hat{\Phi}f = O(\partial f).$$


\begin{thebibliography}{}

\bibitem{landau}
L.~D.~Landau, "On the theory of slow combustion", Acta
Physicochimica URSS {\bf 19}, 77 (1944).

\bibitem{darrieus}
G.~Darrieus, {\it Propagation d'un front de flamme}, Presented at
{\it Le congres de Mecanique Appliquee} (1945)(unpublished).

\bibitem{pelce}
P.~Pelce and P.~Clavin, "Influences of hydrodynamics and diffusion
upon the stability limits of laminar premixed flames", J.~Fluid
Mech. {\bf 124}, 219 (1982).

\bibitem{siv1}
D.~M.~Michelson and G.~I.~Sivashinsky, "Nonlinear analysis of
hydrodynamic instability in laminar flames", Acta Astronaut. {\bf
4}, 1207 (1977).

\bibitem{bychkov2}
V.~V.~Bychkov, S.~M.~Golberg, M.~A.~Liberman, and L.~E.~Eriksson,
"Propagation of curved stationary flames in tubes",
Phys.~Rev.~E{\bf 54}, 3713 (1996).

\bibitem{zel1}
Ya.~B.~Zel'dovich, "An effect stabilizing curved laminar flame
front", Prikl.Mat.Teor.Fiz. {\bf 1}, 102 (1966) (in russian).

\bibitem{zel2}
Ya.~B.~Zel'dovich, G.~I.~Barenblatt, V.~B.~Librovich, and
G.~M.~Makhviladze, {\it The Mathematical Theory of Combustion and
Explosion} (Consultants Bureau, New York, 1985).

\bibitem{siv2}
G.~I.~Sivashinsky, "Nonlinear analysis of hydrodinamic instability
in laminar flames", Acta Astronaut. {\bf 4}, 1177 (1977).

\bibitem{frankel}
M.~L.~Frankel, "An equation of surface dynamics modelling flame
fronts as density discontinuies in potential flow", Phys.~Fluids A
{\bf 2}, 1879 (1990).

\bibitem{bychkov1}
V.~V.~Bychkov, "Nonlinear equation for a curved stationary flame
and the flame velocity", Phys.~Fluids. {\bf 10}, 2091 (1998).

\bibitem{bychkov3}
V.~V.~Bychkov, K.~A.~Kovalev, and M.~A.~Liberman, Phys.~Rev.~E{\bf
60}, 2897 (1999).

\bibitem{matalon}
M.~Matalon and B.~J.~Matkowsky, "Flames as gasdynamic
discontinuities", J.~Fluid Mech. {\bf 124}, 239 (1982).

\bibitem{zhdanov}
S.~K.~Zhdanov and B.~A.~Trubnikov, J.~Exp.~Theor.~Phys. {\bf 68},
65 (1989).

\bibitem{landafshitz}
L.D. Landau, E.M. Lifshitz, {\it Fluid Mechanics} (Pergamon,
Oxford, 1987).

\bibitem{henon}
O.~Thual, U.~Frish, and M.~Henon, "Application of pole
decomposition to an equation governing the dynamics of wrinkled
flames", J.~Phys. (France) {\bf 46}, 1485 (1985).

\bibitem{joulin}
G.~Joulin, "On the Zhdanov-Trubnikov equation for premixed flame
stability", J.~Exp.~Theor.~Phys. {\bf 73}, 234 (1991).

\bibitem{siv3}
M.~Rahibe, N.~Aubry, G.~I.~Sivashinsky, and R.~Lima, "Formation of
wrinkles in outwardly propagating flames", Phys.~Rev.~E{\bf 52},
3675 (1995).

\bibitem{siv4}
M.~Rahibe, N.~Aubry, and G.~I.~Sivashinsky, "Stability of pole
solutions for planar propagating flames", Phys.~Rev.~E{\bf 54},
4958 (1996).

\bibitem{siv5}
M.~Rahibe, N.~Aubry, and G.~I.~Sivashinsky, "Instability of pole
solutions for planar propagating flames in sufficiently large
domains", Combust.~Theory Modelling {\bf 2}, 19 (1998).

\end{thebibliography}
\end{document}